\documentclass[showpacs,superscriptaddress,aps,pre,twocolumn]{revtex4-1}

\usepackage{graphicx}
\usepackage{amsmath}
\usepackage{dcolumn}
\usepackage{bm}
\usepackage{soul}
\usepackage[normalem]{ulem}
\usepackage{color}
\usepackage[caption=false]{subfig}
\usepackage[T1]{fontenc}
\usepackage{float}
\usepackage[12h=true]{scrtime}
\usepackage[dvipsnames]{xcolor}
\usepackage[bottom]{footmisc}
\usepackage{xcolor,lipsum}
\definecolor{mintbg}{rgb}{.63,.79,.95}
\colorlet{lightmintbg}{mintbg!40}
\usepackage{blindtext}

\usepackage{graphicx}
\usepackage[dvipsnames]{xcolor}

\newcommand{\B}[1]{{\bm{#1}}}
\newcommand{\C}[1]{{\mathcal{#1}}}
\begin{document}

\title{Odd Dipole Screening in Radial Inflation}

\author{Yang Fu}
\affiliation{Beijing National Laboratory for Condensed Matter Physics and Laboratory of Soft Matter Physics, Institute of Physics, Chinese Academy of Sciences, Beijing 100190, China}
\author{H. George E. Hentschel}
\affiliation{Department of Physics, Emory University, Atlanta, Georgia-30322, USA}
\author{Pawandeep Kaur}
\affiliation{Department of Chemical \& Biological Physics, Weizmann Institute of Science, Rehovot-7610001, Israel}
\author{Avanish Kumar}
\affiliation{Department of Chemical \& Biological Physics, Weizmann Institute of Science, Rehovot-7610001, Israel}
\author{Itamar Procaccia}
\affiliation{Department of Chemical \& Biological Physics, Weizmann Institute of Science, Rehovot-7610001, Israel}
\affiliation{Sino-Europe Complex Science Center, School of Mathematics, North University of China, Shanxi, Taiyuan 030051, China.}
\date{\today}

\begin{abstract}
The inflation of an inner radial (or spherical) cavity in an amorphous solids confined in a disk (or a sphere), served as a fruitful case model for studying the effects of plastic deformations on the mechanical response. It was shown that when the field associated with Eshelby quadrupolar charges is non-uniform, the displacement field is riddled with dipole charges that screen elasticity, reminiscent of Debye monopoles screening in electrostatics. In this paper we look deeper into the screening phenomenon, taking into account the consequences of irreversibility that are associated with the breaking of Chiral symmetry. We consider the equations for the displacement field with the presence of ``Odd Dipole Screening", solve them analytically and compare with numerical simulations. Suggestions how to test the theory in experiments are provided.  
\end{abstract}
\maketitle
\section{Introduction}

In a series of recent papers the phenomenon of screening in the mechanical response of amorphous solids to non-uniform stresses or strains was examined theoretically\cite{21LMMPRS,23CMP,24KP,24JPS,24HPPS}, experimentally \cite{22MMPRSZ,24CSWDM} and simulationally \cite{22BMP,22KMPS,23MMPR}. The phenomenon at question deals with an amorphous solid that is at mechanical equilibrium (i.e. with vanishing net force on each constituent particle), that is subjected to strain, leading to loss of mechanical equilibrium. Upon relaxation back to mechanical equilibrium, the positions of constituent particles change, until a new configuration is attained in which the net force on each particle vanishes. We are interested in the displacement field $\B d(\B r)$ which is the difference between the final and initial in positions of the centers of mass of each particle. 

For a classical elastic solid the equation obeyed by displacement field $\B {d}$ reads \cite{LandauElasticity}:
\begin{equation}\label{L1}
\Delta \B {d} + (1+\tilde\lambda )\B \nabla (\B \nabla \cdot \B{d}) =0, \quad \tilde\lambda \equiv \lambda/\mu \ .
\end{equation}
Here $\lambda$ and $\mu$ are the classical Lame' coefficients. For an amorphous solids things are different. First, generically any strain on amorphous solids results in plastic deformation \cite{10KLP,11HKLP}. Typical plastic events are quadrupolar in nature \cite{99ML,06ML}, called sometime ``Eshelby inclusions" due to the resemblance to the Eshelby solution of the displacement field caused by forcing a circle to change to an ellipse within an elastic material \cite{54Esh}. When the density of quadrupolar events is finite, one refers to the resulting quadrupolar field as $Q^{\alpha\beta}(\B r)$. It was shown the creation of such a field leads to a renormalization of the elastic moduli \cite{21LMMPRS,23CMP}. Further, when the quadrupolar {\em field} of such events is non-uniform, it was explained that gradients of the field act as effective dipoles,
\begin{equation}
\C	P^\alpha \equiv \partial_\beta Q^{\alpha\beta} \ .
\end{equation}
 When these are present, the equation satisfied by the displacement field change drastically, reading \cite{21LMMPRS,23CMP}
\begin{equation}\label{L2}
 \Delta \B {d} + (1+ \tilde\lambda)\B \nabla (\B \nabla \cdot \B {d}) +\B \Gamma \B {d} =0.
\end{equation}
Here $\B \Gamma$ is a tensor that needs to be specified, containing screening parameters. In previous work it was assumed that in isotropic and homogeneous amorphous solids one can assume that 
$\Gamma$ is diagonal, leading to an equation to be solved of the form  
\begin{equation}\label{L}
 \Delta \B {d} + (1+\tilde\lambda )\B \nabla (\B \nabla \cdot \B {d}) +k^2\B {d} =0.
\end{equation}
The term $k^2\B {d}$ is responsible for translational symmetry breaking, the introduction of a typical length scale $\ell$, 
$\ell \sim k^{-1}$, and to screening phenomena that change dramatically the expected displacement field $\B d$ from the predictions of Eq.~(\ref{L1}). While experiments and simulations provided support for the predictions of Eq.~(\ref{L}), there was one consequence that was found to be in only fair agreement. This is a resulting constitutive equation that relates the dipole and the displacement fields
\begin{equation}
	\B {\C P} =-\kappa^2 \B d\ . 
	\label{Pvsd}
\end{equation}
This constitutive relations means that the dipole and the displacement fields are expected to be co-linear and opposite in direction. This prediction was not accurately corroborated in simulations, an angle was distinctly existing between the prediction and the measured direction of the dipole field \cite{23MMPR}.

The resolution of this issue was offered in Ref.~\cite{24CSWDM}. Eq.~(\ref{L2}) was derived \cite{21LMMPRS,23CMP} from a potential energy, which implies a symmetric tensor $\Gamma^{\alpha\beta} =\Gamma^{\beta\alpha}$. However, in typical amorphous solids energy is not conserved due to plastic rearrangements. Straining along a closed loop in strain space is likely to bring the system to a new state rather than the initial one. This suggests that a diagonal tensor $\B \Gamma$ cannot tell the whole story. To remove the constraint of energy conservation, but keeping a successful screening theory, we allow $\B \Gamma$ to contain an anti-symmetric screening parameter, of the form
\begin{equation}
\B \Gamma=
\begin{bmatrix}
	\kappa_{e}^{2} & \mp\kappa_o^{2}  \\
	\pm\kappa_o^{2} & \kappa_{e}^{2} 
\end{bmatrix} \ .
\label{odd}
\end{equation}
The $\mp$ sign of $\kappa_0^2$ means that it can be negative or positive, with an anti-symmetric counterpart $\pm \kappa_0^2$. The presence of this odd tensor in the theory has led to the term ``odd dipole screening". The existence of the anti-symmetric tensor $\B \Gamma$ leads to breaking Chiral symmetry. Even for purely radial inflation as discussed below, the resulting displacement field exhibits non-trivial transverse (angular) response, as will be shown explicitly in the next Section. 
Denoting the angle formed between the (inverse) direction of the displacement field and the dipole field as $\Phi$, one predicts that 
\begin{equation}
	\tan \Phi = \pm\frac{\kappa_o^2}{\kappa_e^2} \  .
	\label{defphi}
\end{equation} 
The angle $\Phi$ can be positive or negative, depending on the sign of $\kappa_o^2$, but is expected to be the same for all the angles between the dipole and displacement fields in a given realization \cite{24CSWDM}. Tests of this prediction are offered below.

The structure of this paper is as follows: in Sect.~\ref{eqs} we discuss the equations to be solved for two-dimensional radial inflation of an inner disk, and provide their analytic solutions. We then discuss typical numerical simulations using classical glass formers, and compare the numerical measurements to the analytic solutions. A good agreement is reported. In Sect.~\ref{angle} we turn to a deeper analysis of the consequences of Chiral symmetry breaking. In particular we want to verify the prediction embodied in Eq.~(\ref{defphi}). To this aim we will start from the measured displacement field, extract the quadrupolar field associated with the displacement, and from it the dipole field. Having done so we can examine the angle between the dipole field and the inverse direction of the displacement field. The analysis corroborates Eq.~(\ref{defphi}) with the actual measured values of the screening parameters $\kappa_e$ and $\kappa_o$. Section~\ref{summary} offers a summary and discussion of the paper, including suggestions how to test the theory in experiments. 
\begin{figure}
	\includegraphics[width=0.30\textwidth]{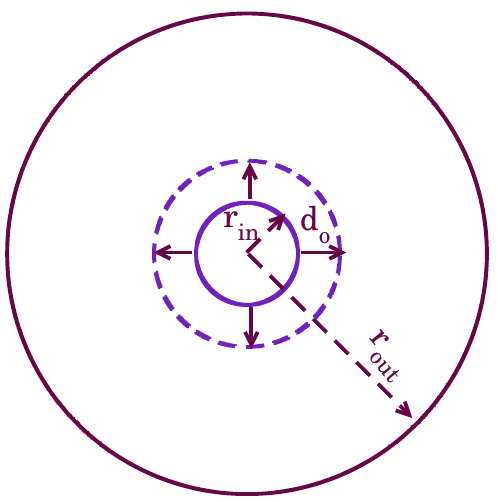}
	\caption{Schematic diagram of the geometry under discussion. Material is confined in an annulus between an outer disk of radius $r_{\rm out}$ and an inner boundary of radius $r_{\rm in}$. The inner radius is then inflated to $r_{\rm in}+d_0$ and then the material is relaxed back to mechanical equilibrium}
	\label{diagram}
\end{figure}
\section{Translational and Chiral Symmetry Breaking}
\label{eqs}

As always, the solution of differential equations depend on the boundary conditions. To make the novel points very clear, we will focus in this paper on two dimensions, having in mind an amorphous solids confined between an outer disk of radius $r_{\rm out}$ and an inner boundary of radius $r_{\rm in}$, cf. Fig.~\ref{diagram}.

After equilibration, the inner circle is inflated, $r_{\rm in}\to r_{\rm in}+d_0$, and the system is equilibrated again. As said, the displacement field studied is the difference between the two configurations after and before the inflation. 
\subsection{Solving for the displacement field}
 
To solve Eq.~(\ref{L2}) with $\B \Gamma$ as defined in Eq.~(\ref{odd}), the displacement field $\B d$ can be separated into radial and transverse components
\begin{equation}\label{L3}
\B {d} = d_{r}(r,\theta)\hat{r} + d_{\theta}(r,\theta)\hat{\theta}.
\end{equation}
For concreteness we will present the solution for the case of $\kappa_o^2$ positive, and a parallel analysis can be easily done for a negative $\kappa_o^2$. We then evaluate the screening term as
\begin{equation}
\B \Gamma \B {d}=
  \begin{bmatrix}
    (\kappa_{e}^{2}d_{r} -\kappa_o^{2}d_{\theta})~\hat{r}  \\
    (\kappa_o^{2}d_{r} + \kappa_{e}^{2}d_{\theta})~\hat{\theta} .
  \end{bmatrix} \ .
\label{product}
\end{equation}
Now equation (\ref{L2}) can be decomposed into following coupled differential equations in $r,$ and $\theta$ (with prime and double prime standing for first or second space derivatives):
\begin{widetext}
\begin{equation}\label{L4}
\frac{(\tilde\lambda +2)}{r^2}\left[ r^2d_{r}^{\prime\prime} + rd_{r}^{\prime}-d_{r}   \right] -\frac{(\tilde\lambda +3)}{r^2}\frac{\partial d_{\theta}}{\partial \theta} + \frac{1}{r^2}\frac{\partial^2 d_{r}}{\partial \theta^2} + \frac{(\tilde\lambda +1)}{r} \frac{\partial^2 d_{\theta}}{\partial r\partial \theta} +\kappa_{e}^{2}d_{r} -\kappa_o^{2}d_{\theta} =0 ,
\end{equation}
\begin{equation}\label{L5}
\frac{1}{r^2}\left[ r^2d_{\theta}^{\prime\prime} + rd_{\theta}^{\prime} + \frac{\partial^2 d_{\theta}}{\partial\theta^2}-d_{\theta}   \right] + \frac{(1 +\tilde\lambda)}{r^2}\left[ \frac{\partial^2 d_{\theta}}{\partial\theta^2} + r\frac{\partial^2 d_{r}}{\partial r\partial \theta} \right] + \frac{(\tilde\lambda +3)}{r^2}\frac{\partial d_{r}}{\partial \theta} + \kappa_o^{2}d_{r} + \kappa_{e}^{2}d_{\theta} =0 .
\end{equation}
\end{widetext}
\begin{figure}
	\includegraphics[width=0.375\textwidth]{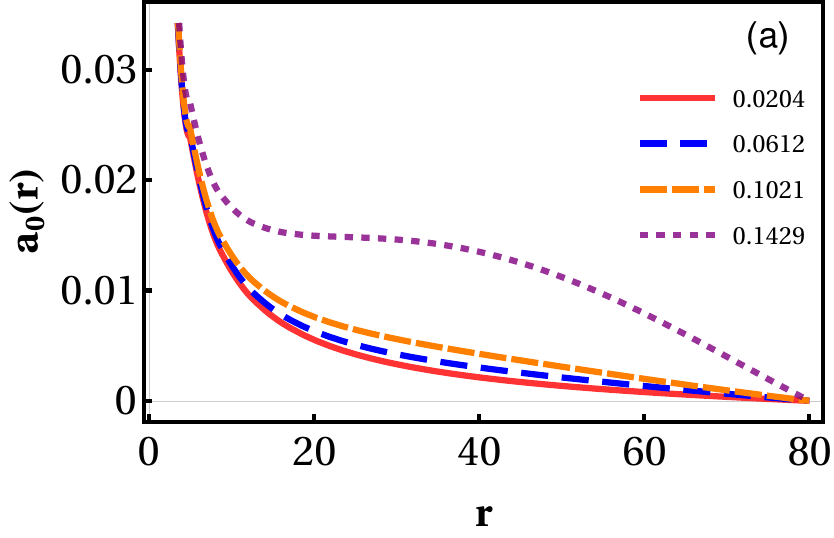}
	\includegraphics[width=0.375\textwidth]{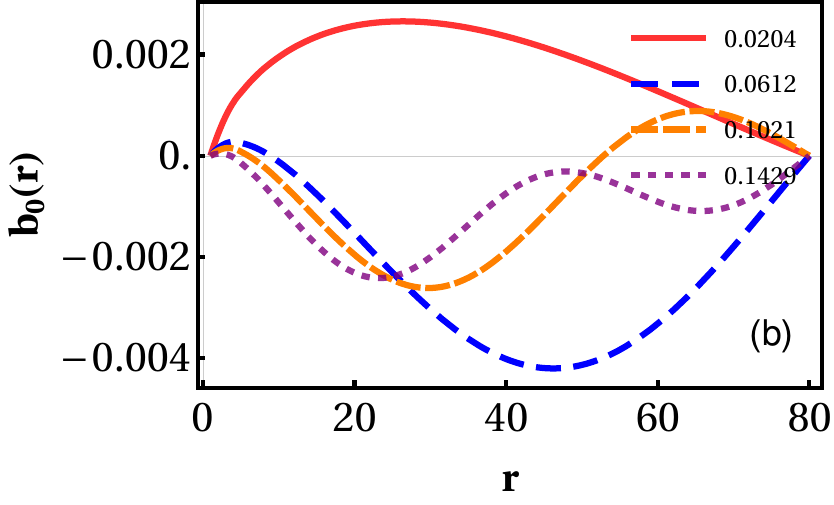}
	\includegraphics[width=0.375\textwidth]{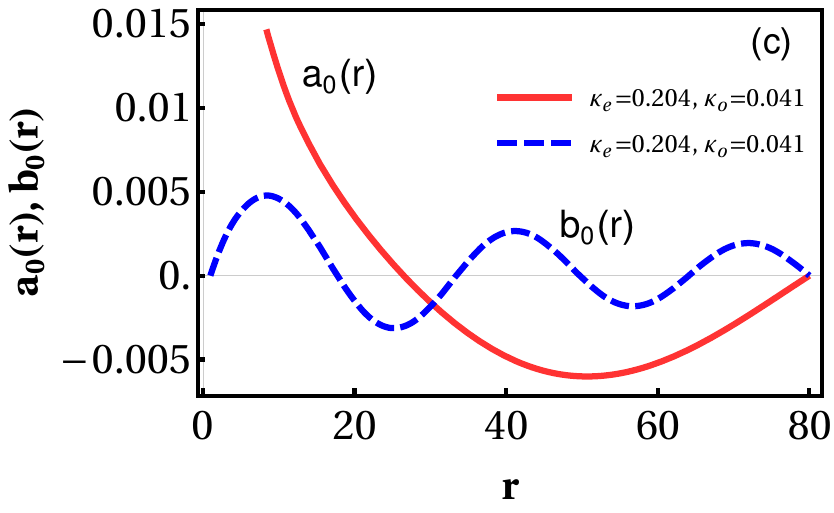}
	\caption{Panel (a) and (b): Typical solutions of $a_0(r)$ and $b_0(r)$ for a fixed
		$\kappa_{\rm o}$=0.041 and for different values of $\kappa_{\rm e}=0.0204,0.0612,0.102,0.1429$. Panel (c): Comparison of $a_0$ and $b_0$ for $\kappa_{\rm o}=0.041$ and  $\kappa_{\rm e}=0.204$. In all these cases the boundary conditions are $a_0(r_{\rm in})=0.1$, $a_0(r_{\rm out})=0$, $b_0(r_{\rm in})=0$,$b_0(r_{\rm out})=0$.} 
	\label{math1}
\end{figure}
\begin{figure}
	\includegraphics[width=0.375\textwidth]{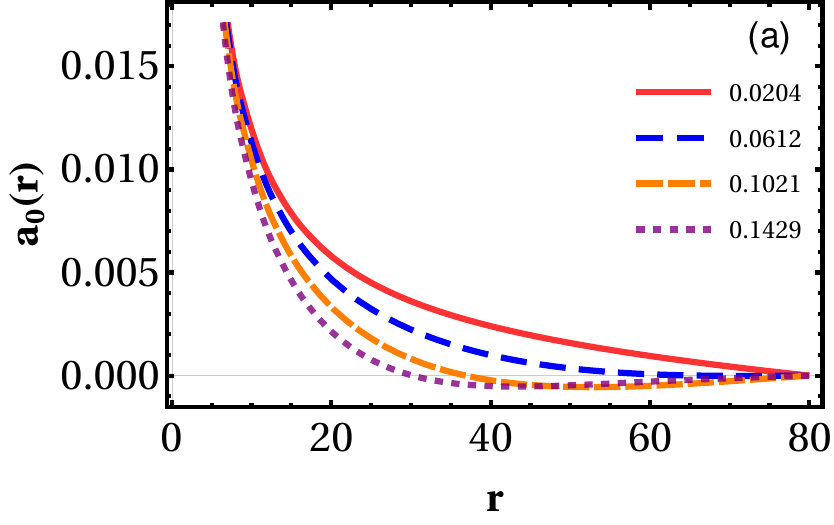}
	\includegraphics[width=0.375\textwidth]{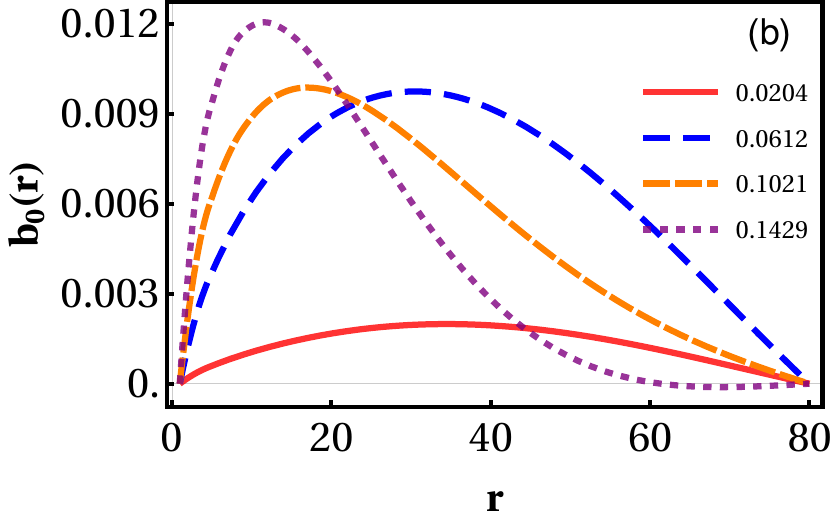}
	\includegraphics[width=0.375\textwidth]{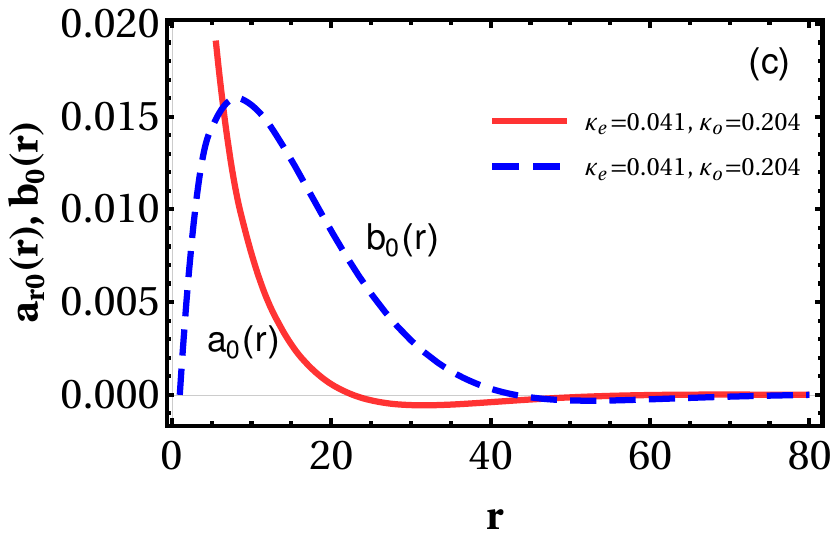}
	\caption{Panel (a) and (b): Typical solutions of $a_0(r)$ and $b_0(r)$ for a fixed
		$\kappa_{\rm e}$=0.041 and for different values of $\kappa_{\rm o}=0.0204,0.0612,0.102,0.1429$. Panel (c): Comparison of $a_0$ and $b_0$ for $\kappa_{\rm o}=0.204$ and  $\kappa_{\rm e}=0.041$. The boundary conditions are the same as in Figs.~\ref{math1}}.
	\label{math2}
\end{figure}

Taking into account the periodic boundary conditions on the angle $\theta$ we can now seek a solution of these equations
in the form of Fourier series for $d_{r}(r,\theta)$ and $d_{\theta}(r,\theta)$ respectively:
\begin{align}\label{L6}
&d_{r}(r,\theta) = a_{0}(r) + \sum_{n=1}^{\infty}\left[  a_{n}(r)\cos(n\theta) + c_{n}(r)\sin(n\theta)\right] , \nonumber \\
&d_{\theta}(r,\theta)  = b_{0}(r) + \sum_{n=1}^{\infty} \left[ g_{n}(r)\cos(n\theta) +e_{n}(r)\sin(n\theta)\right] .
\end{align}
Of course, the more coefficients we keep, the more equations we need to solve. However, due to the orthogonality of the Fourier coefficients and the linearity of the equations,  different order coefficients do not mix. Thus, for simplicity, to demonstrate the decoupling,  we consider only the first-order Fourier terms, $n=1$. We have
\begin{align}\label{L7}
&d_{r}(r,\theta) = a_{0}(r) + a_{1}(r)\cos(\theta) + c_{1}(r)\sin(\theta), \nonumber \\
&d_{\theta}(r,\theta)  = b_{0}(r) + g_{1}(r)\cos(\theta) + e_{1}(r)\sin(\theta).
\end{align}

After substitution of this ansatz, and after some simplifications, Eqs.~(\ref{L5}) and (\ref{L6}) respectively take the following forms:
\begin{widetext}

\begin{align}\label{L10}
&\left[ r^2a_{0}^{\prime\prime}(r) + ra_{0}^{\prime}(r)-a_{0}(r)   \right] +  \frac{\kappa_{e}^{2}r^2}{(\tilde\lambda +2)}  a_{0}(r) -\frac{\kappa_{0}^{2}r^2}{(\tilde\lambda +2)}  b_{0}(r) \nonumber \\
& + \left[  \left[ r^2a_{1}^{\prime\prime}(r) + ra_{1}^{\prime}(r)   \right] +  \frac{(\kappa_{e}^{2}r^2 -\tilde\lambda -3)}{(\tilde\lambda +2)} a_{1}(r) - \frac{\kappa_o^{2}r^2}{(\tilde\lambda +2)} g_{1}(r) + \frac{(\tilde\lambda+1)r}{(\tilde\lambda +2)} e_{1}^{\prime}(r) -\frac{(\tilde\lambda +3)}{(\tilde\lambda +2)}e_{1}(r)   \right] \cos(\theta) \nonumber \\
& + \left[ \left[ r^2c_{1}^{\prime\prime}(r) + rc_{1}^{\prime}(r)   \right] + \frac{(\kappa_{e}^{2} r^2 -\tilde\lambda-3)}{(\tilde\lambda +2)} c_{1}(r) - \frac{\kappa_o^{2} r^2}{(\tilde\lambda +2)} e_{1}(r)- \frac{(\tilde\lambda+1)r}{(\tilde\lambda +2)} g_{1}^{\prime}(r) + \frac{(\tilde\lambda +3)}{(\tilde\lambda +2)}g_{1}(r)   \right] \sin(\theta) 
=0,  
\end{align}
\begin{align}\label{L11}
&\left[ r^2b_{0}^{\prime\prime}(r) + rb_{0}^{\prime}(r)-b_{0}(r)   \right] + \kappa_{e}^{2} r^2 b_{0}(r) +   \kappa_{0}^{2} r^{2}  a_{0}(r) \nonumber \\
& + \left[  \left[ r^2g_{1}^{\prime\prime}(r) + rg_{1}^{\prime}(r)  \right]  + (\kappa_{e}^{2} r^2 -\tilde\lambda-3) g_{1}(r) +  \kappa_o^{2}r^2 a_{1}(r) + (\tilde\lambda +1)r c_{1}^{\prime}(r) +(\tilde\lambda+3) c_{1}(r)  \right] \cos(\theta) \nonumber \\
& + \left[  \left[ r^2e_{1}^{\prime\prime}(r) + re_{1}^{\prime}(r)  \right]   + (\kappa_{e}^{2}r^2-\tilde\lambda-3) e_{1}(r) +  \kappa_o^{2}r^2 c_{1}(r) -(\tilde\lambda+1)r a_{1}^{\prime}(r) -(\tilde\lambda +3) a_{1}(r)  \right] \sin(\theta)  =0. 
\end{align}
\end{widetext}
Since each line in the above equations has to vanish separately, these two equations produce a system of six coupled differential equations for the coefficients $a_0, b_0, a_1, b_1, c_1$ and $d_1$.
It is important to realize that keeping higher order terms in Eq.~(\ref{L7}) would not change these equations, but will only add more independent equations for higher order coefficients. 
\begin{figure}
	\includegraphics[width=0.35\textwidth]{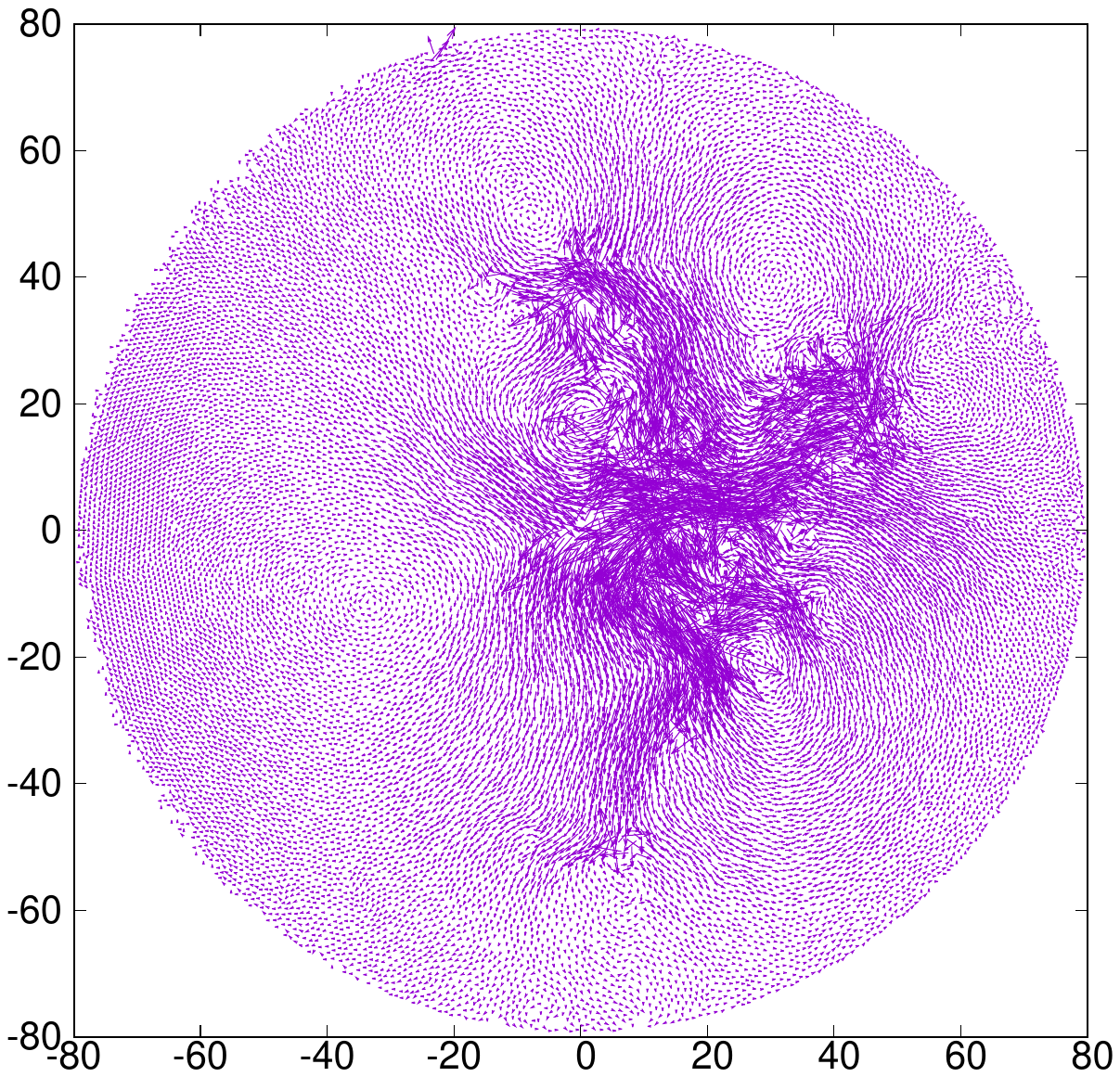}
	\includegraphics[width=0.35\textwidth]{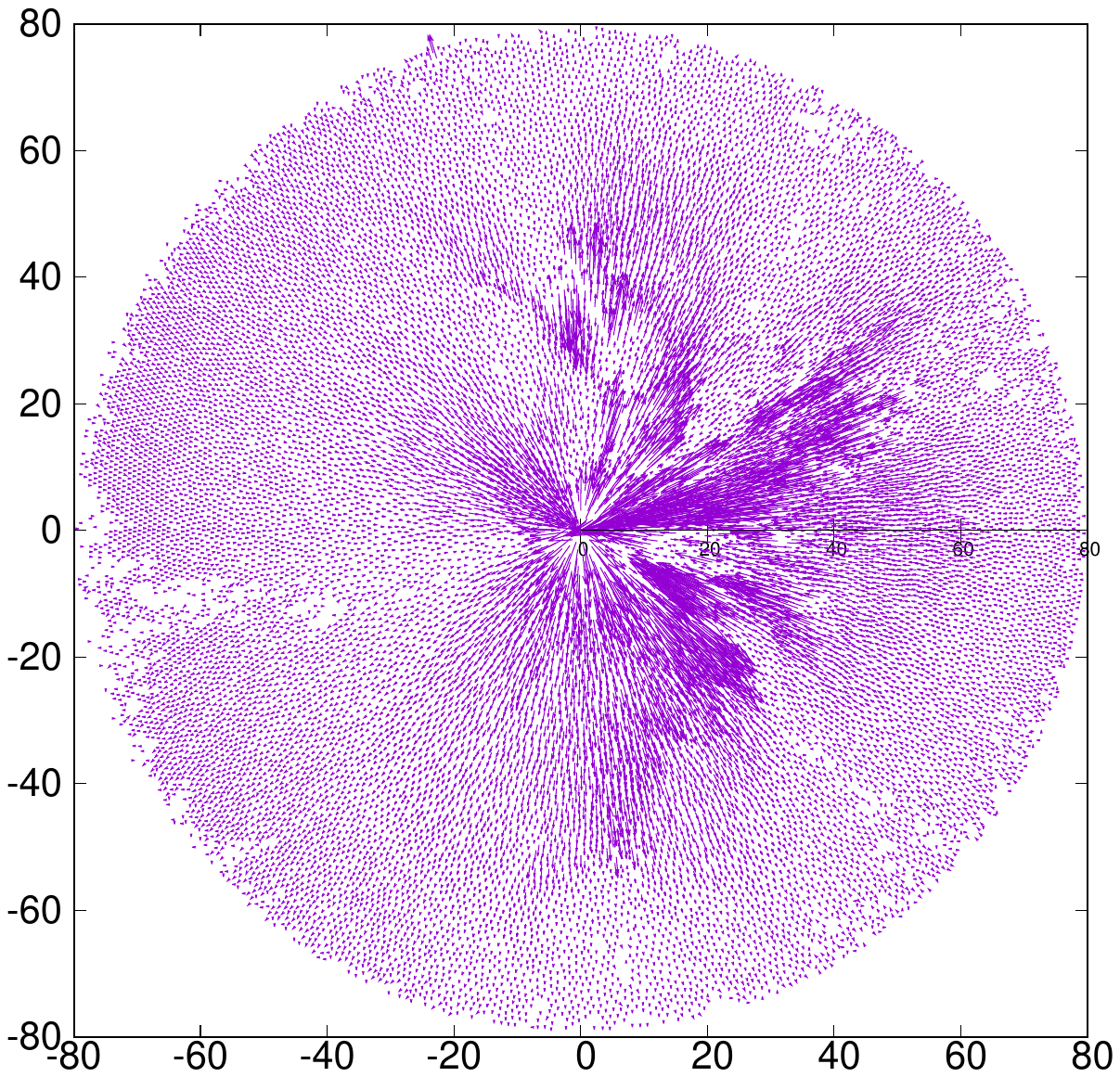}
	\includegraphics[width=0.35\textwidth]{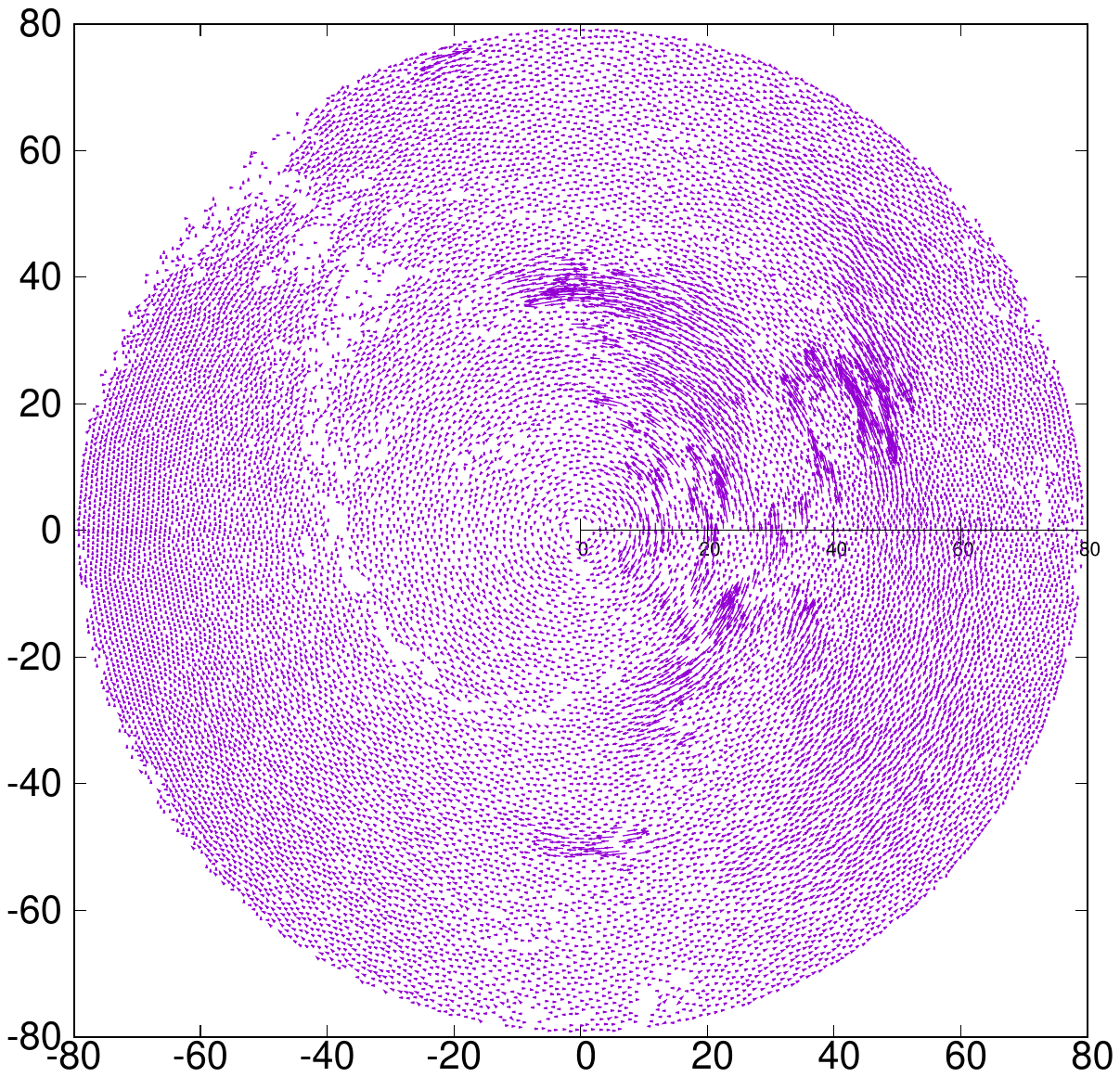}
	\caption{Panel (a): A typical displacement field resulting from an inflation of the inner boundary. Panel (b): the radial
		component of the same displacement field. Panel (c): the transverse component of the same field.}
	\label{disp}
\end{figure}

\subsection{Analytic solutions} 

For the case of radial inflation these equations simplify further, since the boundary conditions are  $a_{0}(r=r_{\rm in})=d_{0}$ and all the other coefficients are zero at $r_{\rm in}$. At $r_{\rm out}$ all the coefficients must vanish. Thus the only remaining nonzero coefficients are $a_0(r)$ and $b_0(r)$. All the higher Fourier modes will have zero values unless we provide different boundary conditions.

The equations for $a_0$ and $b_0$ can be solved analytically, as is demonstrated in Appendix \ref{anal}. The final solutions are
shown in Eqs.~(\ref{good1}) and (\ref{good2}).
Typical solutions are shown in Figs.~\ref{math1} and \ref{math2}. Here $\tilde \lambda=9.17$. We note that in the limit $\tilde \lambda\to \infty$ the analytic solutions converge the the purely elastic solution.

To compare with numerical simulations, we need to extract data for the coefficients $a_0(r)$ and $b_0(r)$ from the measured displacement field. To this aim we compute the angle averages
\begin{eqnarray}\label{L14}
&& a_{0}(r) = \frac{1}{2\pi}\oint_{0}^{2\pi} d_{r}(r,\theta) d\theta, \\
&& b_{0}(r) = \frac{1}{2\pi}\oint_{0}^{2\pi} d_{\theta}(r,\theta) d\theta, \nonumber 
\end{eqnarray}
\subsection{Comparison with simulations}

The description of the numerical procedures for the preparation of our amorphous solids for the inflation of the inner disk is offered in Appendix \ref{numerics}. In all the simulations below the inflation magnitude $d_0$ is between 10\% to 20\% of $r_{\rm in}$.

A typical displacement field and its radial and transverse components are shown in Fig.~\ref{disp}.
Next we compare to the analytic solutions using Eqs.~(\ref{L14}).

When we compare the results of the simulations to the exact solutions of the type displayed in  Figs.~\ref{math1} and \ref{math2}, we run into the usual difficulties encountered when one compares continuum theory to granular simulations. While one expects excellent agreement in the bulk of the system, on the scale of the grains one can have deviations. In particular the boundary conditions at $r_{\rm in}$ can be an issue. While we find that the radial boundary condition $a_{0}(r=r_{\rm in})=d_{0}$ is faithfully obeyed, the other boundary condition, i.e.  $b_{0}(r=r_{\rm in})=0$ is not obeyed precisely.  Even though the inflation is purely radial, due to the disorder and the interaction of the inner boundary with the bulk, the close particles touching the boundary receive some transverse displacement. Thus, in order to compare to the theory, we {\em measured} the transverse displacement at the first layer of disks, and used this as a boundary condition in the analytic solution. We note that having a transverse displacement will excite also higher order Fourier modes, but due to the orthogonality of these modes, the solutions for $a_0$ and $b_0$ will not be affected. Needless to say, the observed displacement field will look more complex when more Fourier modes are present. 

\begin{figure}
	\includegraphics[width=0.35\textwidth]{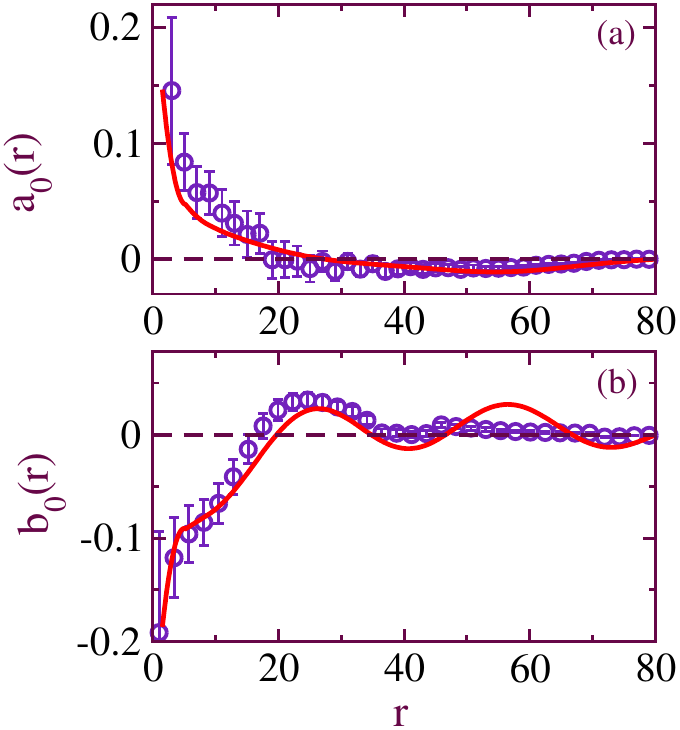}
		\includegraphics[width=0.35\textwidth]{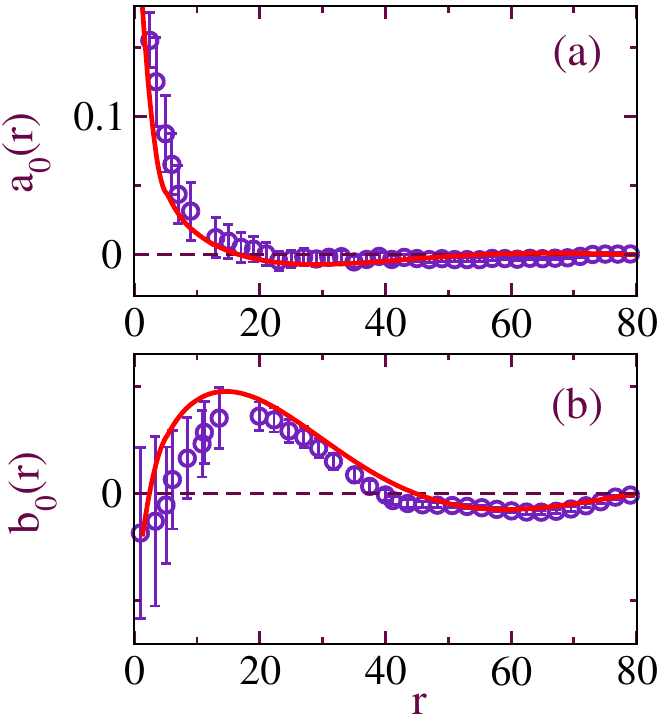}
		\includegraphics[width=0.35\textwidth]{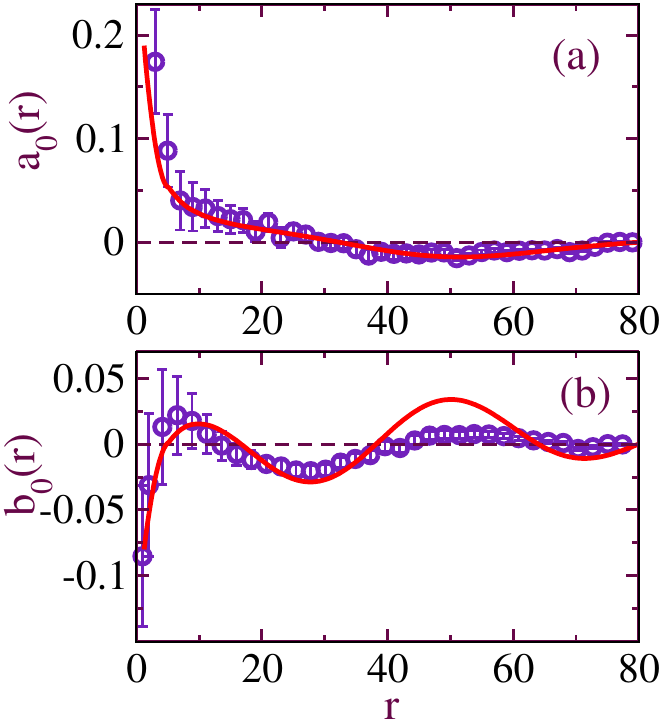}
	\caption{Comparisons of angle averaged radial ($a_0(r)$) and transverse ($b_0(r)$) displacement field with that of analytical solutions (solid line). Upper figure: $r_{in} = 1.25$, $r_{out} = 80$, $\tilde\lambda = 20.09$, $\kappa_e = 0.209$ and $\kappa_o =0.206$. Here $a_0(r_{in}) = 0.16$ and $b_0(r_{in}) = -0.2$. Middle figure: $r_{in} = 1.3$, $r_{out} = 80$,  $\tilde\lambda = 7.9$, $\kappa_e = 0.099$ and $\kappa_o =0.149$. Here $a_0(r_{in}) = 0.18$ and $b_0(r_{in}) = -0.02$. Lower figure: $r_{in} = 1.25$, $r_{out} = 80$, $\tilde\lambda = 10.46$, $\kappa_e = 0.165$ and $\kappa_o =0.155$. Here $a_0(r_{in}) = 0.19$ and $b_0(r_{in}) = -0.08$.}
\label{compare}
\end{figure}

Typical comparisons of numerical and exact solution are shown in Fig.~\ref{compare}.
Here are three independent realizations, with different material parameters, and different geometries.
We note that the analytic solutions depend on the actual elastic moduli $\mu$ and $\lambda$. We have computed these for the actual realizations after equilibration, from the microscopic theory that is described for example in Ref. \cite{89Lutsko}. In contrast, the screening parameters $\kappa_e$ and $\kappa_o$ were fitted to the data, and were the only free parameters. We find the agreement between data and theory quite satisfactory, despite some slight deviation mainly in $b_0(r)$ at higher values of $r$. We therefore turn now to an even more sensitive test for the fit of the screening parameters, i.e. the angle $\Phi$ defined in Eq.~(\ref{defphi}).

\section{The Consequences of Chiral Symmetry Breaking}
\label{angle} 

In this section we explore the angle between the (inverse) direction of the displacement field and the dipole field. This angle is a direct consequence of the breaking of Chiral symmetry and it must vanish if $\kappa_o=0$, i.e. for a symmetric tensor $\B \Gamma$. To this aim we need to extract the dipole field and find the angle between it and the inverse displacement direction. 

\subsection{Extracting the Quadrupolar and Dipole Fields from displacement data}
To accomplish our aim we must extract first the quadrupolar field associated with the displacement field. Since a quadrupole is represented by the traceless part of the non-affine strain tensor, so we first compute the total strain tensor. We follow the procedure advised in Ref.~\cite{08GHBF}, using the discrete deformation gradient $F$ defined as,
\begin{align}\label{S1}
\varepsilon=  \frac{1}{2}\left[  I - \left( F^{\text{T}}F  \right)^{-1}  \right],  
\end{align}
where $F\equiv \frac{\partial \chi}{\partial \bf{X}} =  \frac{\partial x}{\partial \bf{X}}$ is a second rank tensor. Here $\chi$ maps a particle position from the initial undeformed configuration $\Omega_0$ to final deformed configuration $\Omega_{D}$ defined by $\bf{x = \chi (X)}$.
Here $\bf{X}$ and $\bf{x}$ are the positions of the particles before and after the deformation in the two configurations.

To calculate $F$ in the atomistic simulation we measure the relative displacements between two particles. Before we proceed, we would like to mention that the particle positions have been mapped to grid points. To this aim we have used the Matlab library that uses cubic spline and Linear Interpolation standard methods. Therefore, in all our subsequent discussion particle labels and grid points are synonyms. Let  $\Delta X^{ij} = X^{j} - X^{i}$, and $\Delta x^{ij} = x^{j} - x^{i}$ be the relative displacements in $\Omega_0$ and $\Omega_{D}$ respectively between the particles $i$ and $j$. The transformation from configuration $\Omega_0$ to $\Omega_{D}$ is defined by deformation gradient $F$, which linearly maps the relative \textit{position vector} $\bf{\Delta X^{ij}}$ to $\bf{\Delta x^{ij}}$ as follows,
\begin{align}\label{S2}
\Delta x^{ij} = F^{i}\Delta X^{ij},
\end{align}
where $F^{i}$ is the deformation gradient measured at the particle $i$. Note that, instead of $F$, we use a local deformation gradient $\hat{F}$, which is given by \cite{08GHBF}, 
\begin{equation}\label{S3}
\hat{F}D = A ~~ \text{or} ~~\hat{F}  = AD^{-1},
\end{equation}
where $A$ and $D$ are $2\times 2$ matrices, which are obtained from
\begin{align}\label{S4}
 {A = \sum_{j=1}^{L} \Delta x^{ij}\Delta {X^{ij}}^{\text{T}} w_{j} }, ~ {D = \sum_{j=1}^{L} \Delta X^{ij}\Delta {X^{ij}}^{\text{T}} w_{j} } .
\end{align}
Here $w_{j}$ is the weight function assigned to particle $j$ depending upon its relative distance with the central atom $i$, and $L$ denotes the total neighbors around particle $i$ \cite{08GHBF}. 

The non-affine strain $\varepsilon^{q}$ is obtained after subtracting the affine strain components $\varepsilon^{a}$ from the total strain $\varepsilon$ measured in Eqn.(\ref{S1}),
\begin{align}\label{S5}
 \varepsilon_{\alpha\beta}^{q}  = \varepsilon_{\alpha\beta} - \varepsilon_{\alpha\beta}^{a} = \begin{bmatrix} \varepsilon_{11}^{q} & \varepsilon_{12}^{q}  \\  \varepsilon_{21}^{q} & \varepsilon_{22}^{q} \end{bmatrix}.  
\end{align}
Here $\varepsilon_{\alpha\beta}^{a} = 0.5(\nabla_{\alpha} d^a_{\beta} + \nabla_{\beta}d^a_{\alpha})$, and $\alpha, \beta=1,2$ which are same as $(x,y)$. Note that $\varepsilon_{\alpha\beta}^{q} $ is calculated at each particle label or each grid point. Here, $d^a$ is the quasi-elastic displacement field in purely radial inflation which is given by \cite{21LMMPRS}:
\begin{equation}\label{S6}
d^a(r) = d_o \left[ \frac{r^2-r_{out}^2}{r_{in}^2 -r_{out}^2} \right]  \left( \frac{r_{in}}{r}    \right) .
\end{equation}
\begin{figure}
	\includegraphics[width=0.45\textwidth]{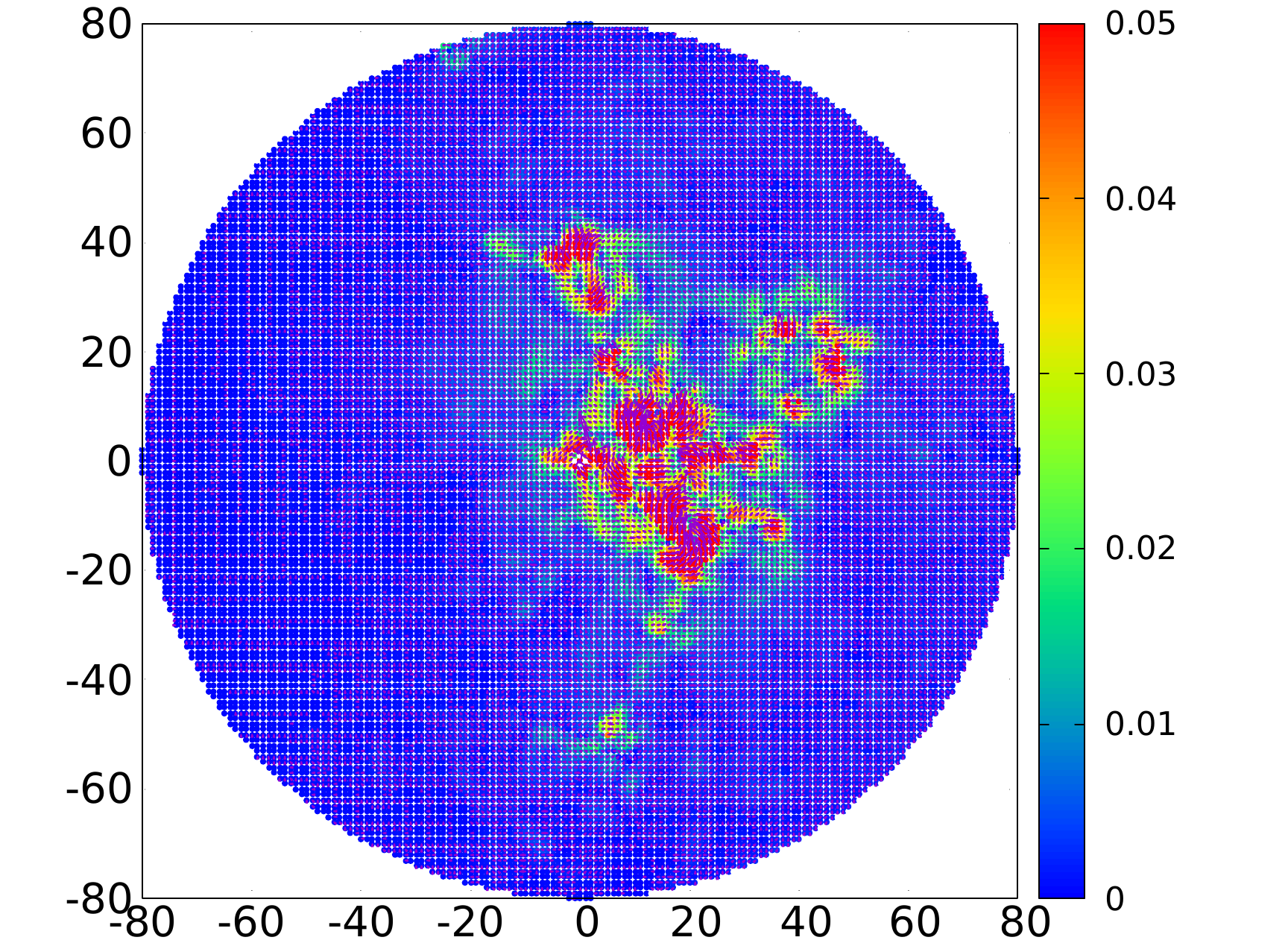}  
	\includegraphics[width=0.45\textwidth]{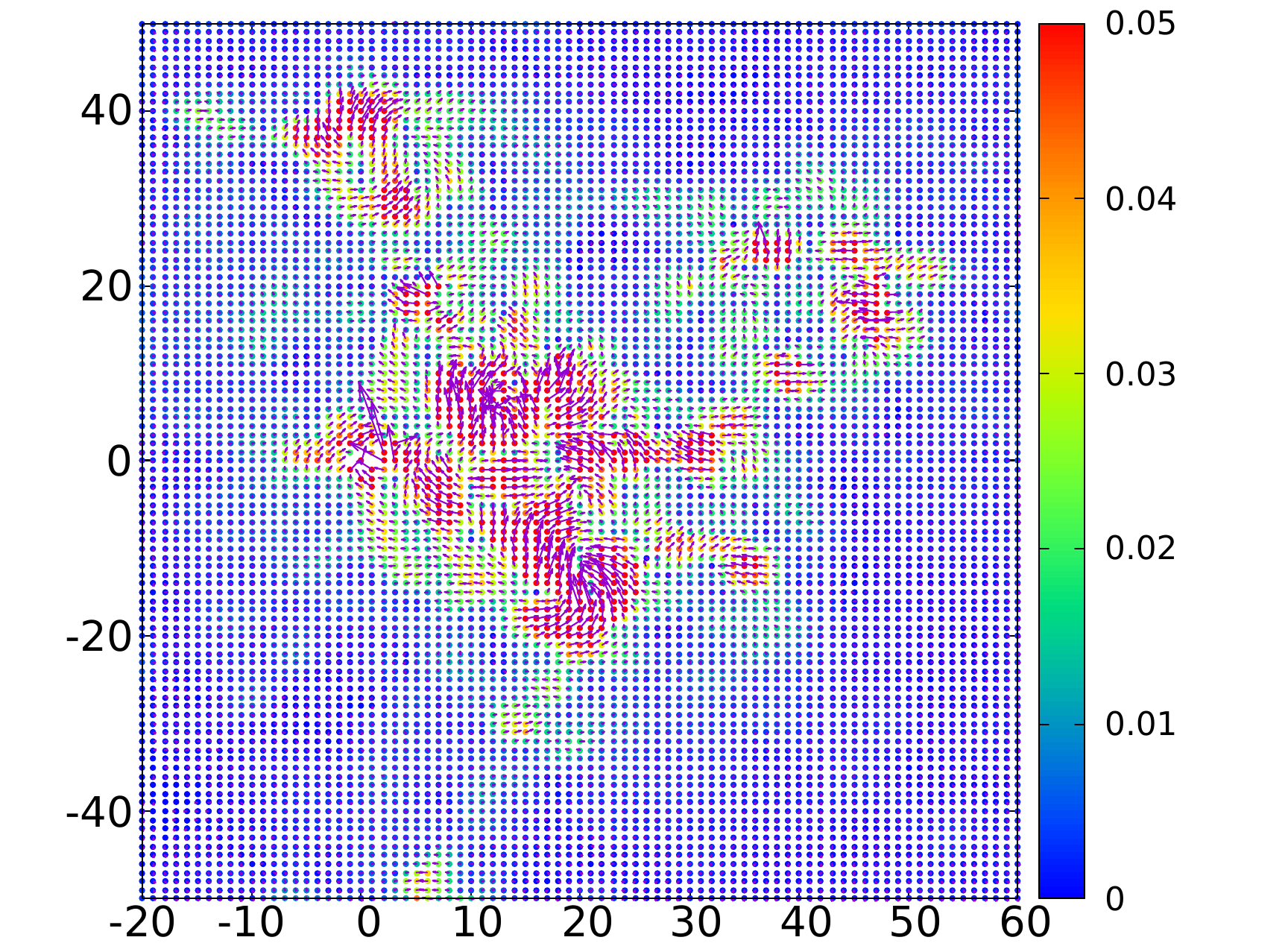}
	\caption{Heat map of the quadrupolar field in typical radial inflation. The dense regions indicate high values of $Q$, and less dense regions correspond to the low values of $Q$ as obtained from Eq.~(\ref{S8}). The arrows represent the principal axis or the directions of the $Q$ field, cf. Eq.~(\ref{S10}).
		The parameters here correspond to the first pair of panels in Fig.~\ref{compare}.
		In the top panel we represent the full quadrupolar field, whereas in the bottom panel we show the magnified view of the most dense region.}
	\label{Quadrupole}
\end{figure}
The non-affine strain $\varepsilon^{q}$ can be decomposed into its trace and its traceless components \cite{23MMPR}:
\begin{align}\label{S7}
\varepsilon^{q} &= \frac{1}{2}\begin{bmatrix} \varepsilon_{11}^{q} + \varepsilon_{22}^{q} &0\\  0&\varepsilon_{11}^{q} + \varepsilon_{22}^{q} \end{bmatrix} +\begin{bmatrix} \varepsilon_{11}^{ts}& \varepsilon_{12}^{ts}\\ \varepsilon_{21}^{ts}  & \varepsilon_{22}^{ts} \end{bmatrix} \ .
\end{align}
The traceless component has a quadrupolar structure, and Eq.~(\ref{S7}) can also be written in the following form,
\begin{align}\label{S9}
	\varepsilon^{q} 
	&=\frac{1}{2}\begin{bmatrix} \varepsilon_{11}^{q} + \varepsilon_{22}^{q} &0\\  0&\varepsilon_{11}^{q} + \varepsilon_{22}^{q} \end{bmatrix} +Q\begin{bmatrix} \cos(2\Theta)& \sin(2\Theta)\\  \sin(2\Theta)&-\cos(2\Theta) \end{bmatrix}.
\end{align}
 The quantity $Q$ is the quadrupolar charge associated with the quadrupolar tensor and is defined as,
\begin{equation}\label{S8}
Q^2 = (\varepsilon_{11}^{ts})^2 + (\varepsilon_{12}^{ts})^2.
\end{equation}

Therefore the orientation of the quadrupolar field is given by,
\begin{equation}\label{S10}
\Theta = \frac{1}{2} \arctan\left( \frac{\varepsilon_{12}^{ts}}{\varepsilon_{11}^{ts}}\right).
\end{equation}
$\Theta$ defines the principal axis of the quadrupolar field. We show a typical  quadrupolar field in Fig.~\ref{Quadrupole}. The data used to generate this and the next figure correspond to the top system in Fig.~\ref{compare}. 

\begin{figure}
	\includegraphics[width=0.405\textwidth]{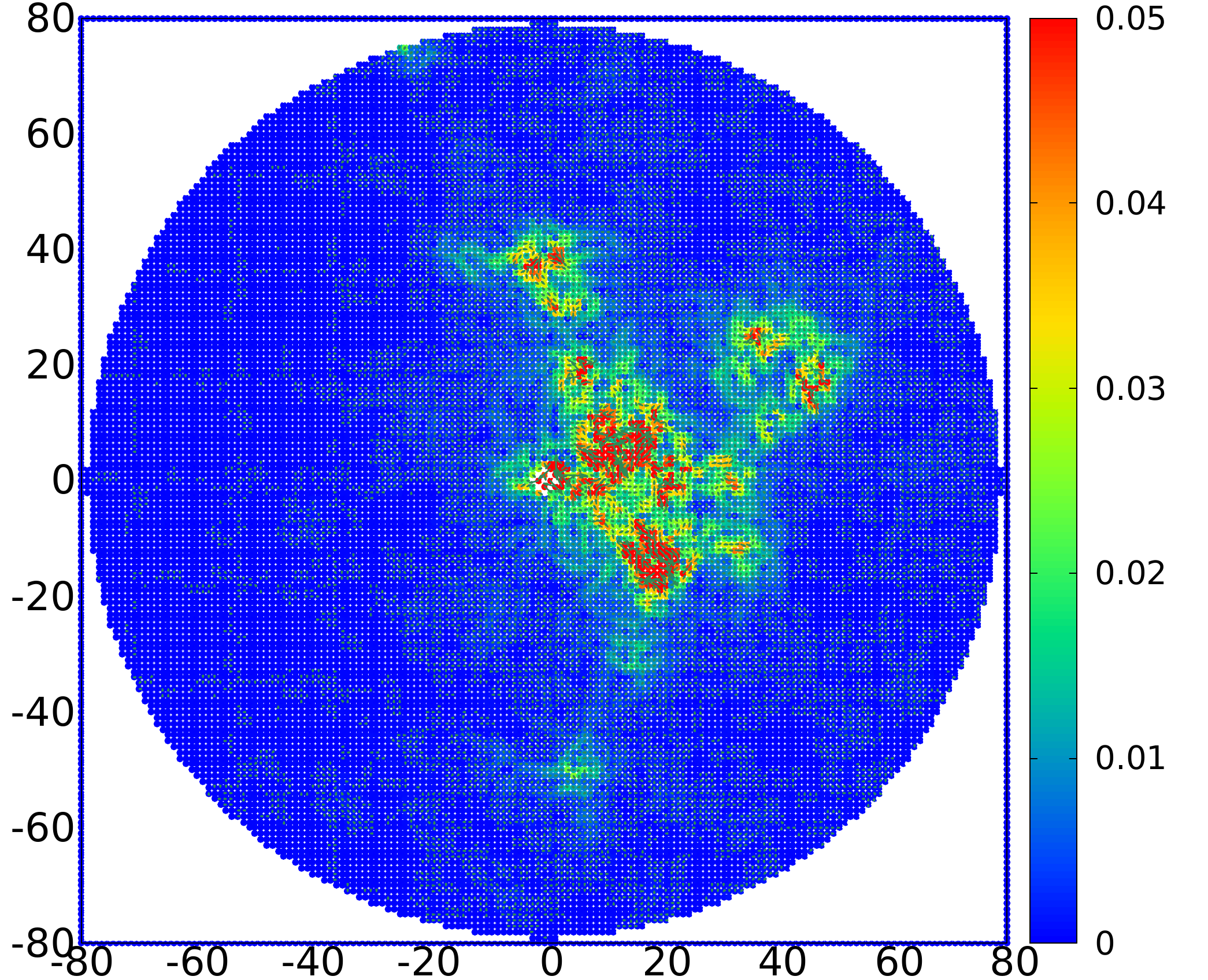}
	\includegraphics[width=0.405\textwidth]{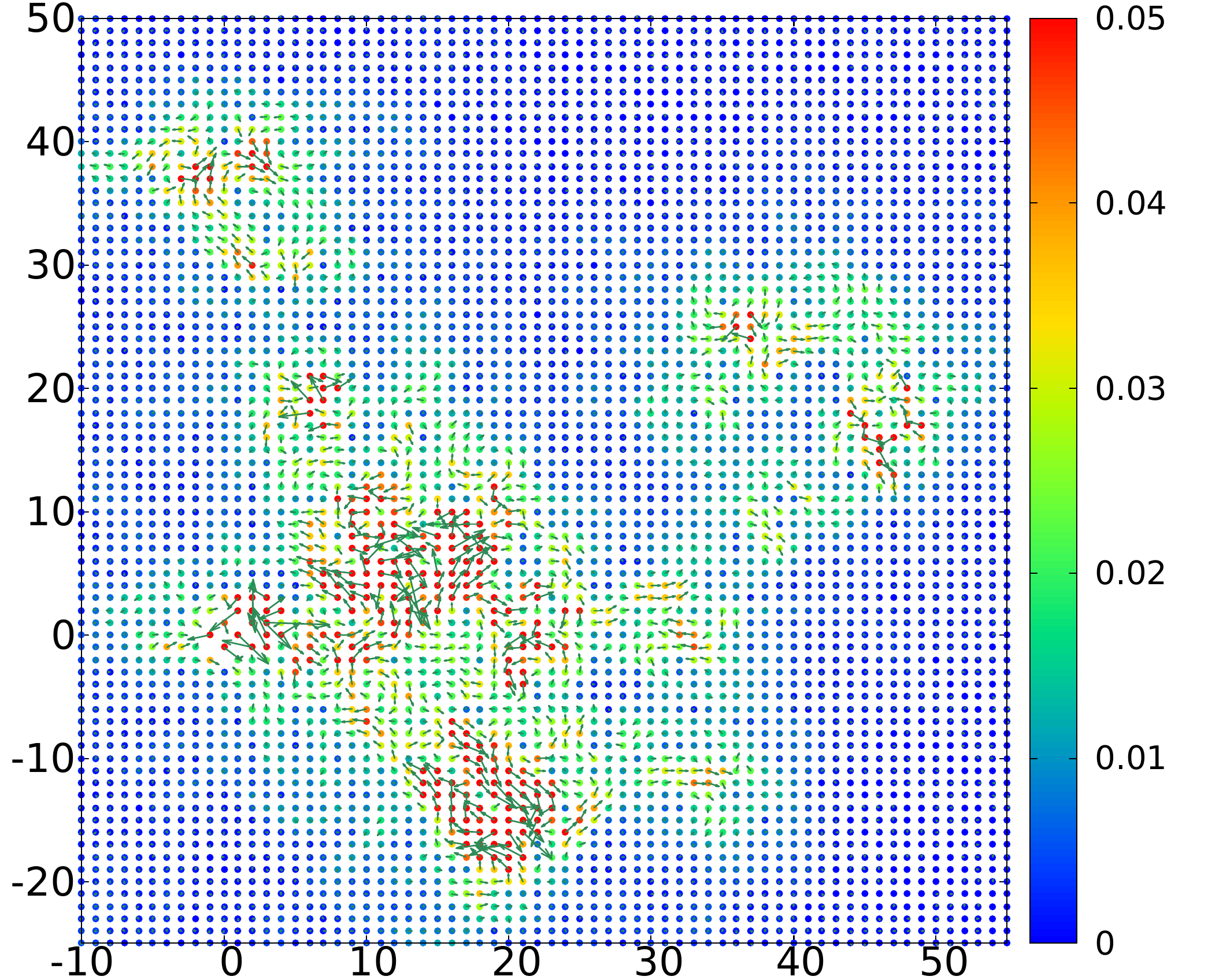}
	\caption{Heat map of the dipole field after a typical radial inflation, obtained by taking the gradient of the quadrupolar field $\C	P^\alpha \equiv \partial_\beta Q^{\alpha\beta} \ $. The dense regions indicate high values of $\C P \ $, and less dense regions correspond to the low values of dipole field $\C P \ $.
The arrows represent the local direction of the dipole field.	 
	 The parameters here correspond to the first pair of panels in Fig.~\ref{compare}.
	 In the top panel we represent the full dipole field, whereas in the bottom panel we show the magnified view of the most dense region.}
	\label{Dipole}
	\end{figure}

Next we calculate the dipole field using the gradient of the quadrupolar field, $\C	P^\alpha \equiv \partial_\beta Q^{\alpha\beta} \ $ at each grid point. The dipole field associated with the quadrupolar field in Fig.~\ref{Quadrupole} is shown  in figure \ref{Dipole}. 
\begin{figure}
	\includegraphics[width=0.43\textwidth]{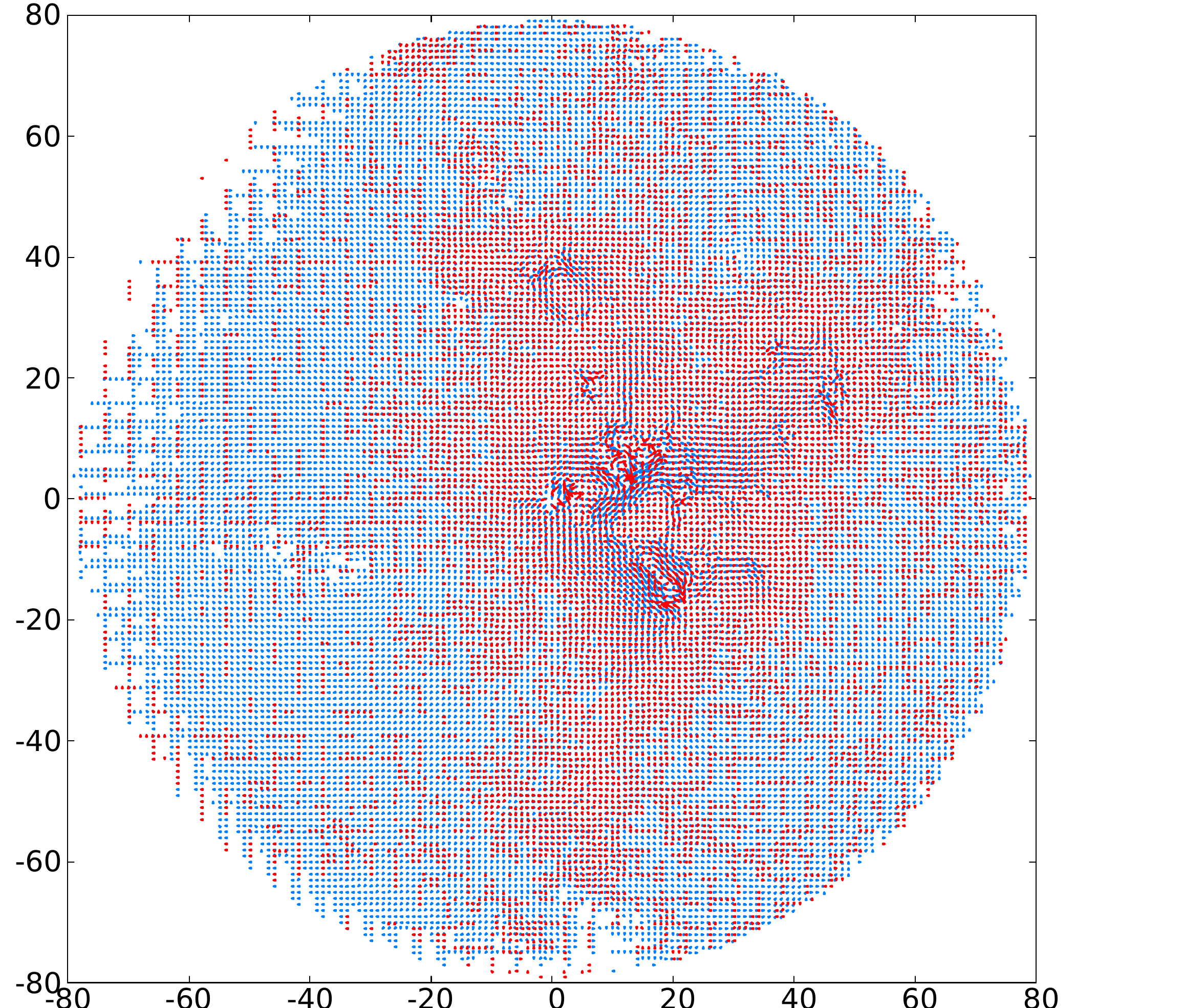}
	\includegraphics[width=0.43\textwidth]{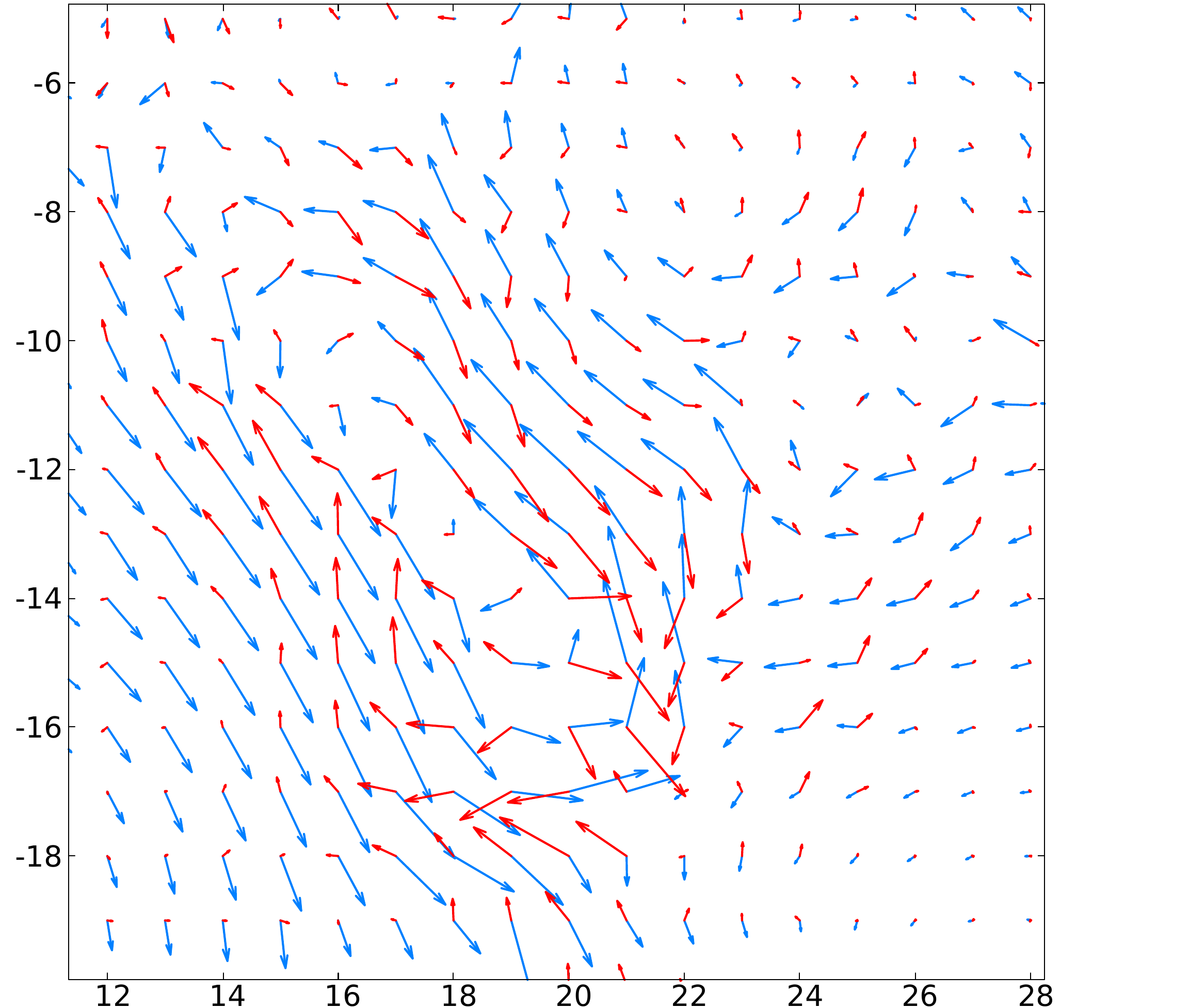}
	\caption{The dipole and displacement vector fields shown together, red arrows for the dipole field and blue arrows for displacement. The parameters here are those of Fig.~\ref{Dipole}.  In the top panel we represent the full dipole and displacement fields, whereas in the bottom panel we show the magnified view of the most dense region. The averaged angle between the dipole and the inverse displacement field from the plots is $46.48$ degrees, whereas theoretically we obtain $\theta = \arctan\left( \frac{\kappa_o^2}{\kappa_{e}^2}\right)  =  44.11$ degrees.}
	\label{Dip-dis}
	\end{figure}
Having the dipole field we compute the angle between the inverse displacement and the dipole field, which is predicted to be the angle $\Phi$ of Eq.~(\ref{defphi}).
In Fig.~\ref{Dip-dis} we show simultaneously the dipole and displacement fields, in two colors, from which we can determine the angles between them. For the present example, we found from simulations an average angle of $46.48^o\pm 0.55^o$ as compared
to an angle of $44.11^o$ from the direct calculation of Eq.~(\ref{defphi}). This close correspondence is typical and similar results were obtained for other simulations of radial inflation.

\section{Summary and discussion}
\label{summary}

It is important to distinguish the present phenomena from those described by Cosserat Elaticity \cite{19RHL} and Odd Elasticity \cite{20SSBSIV}. Both these theories enlarge classical elasticity, that contains only shear and bulk moduli, to describe more general situations. The former theory allows the presence of internal rotational degrees of freedom; In an isotropic Cosserat solid there are six elastic constants rather than two. The latter theory allows active constituents that can propel themselves, resulting in a non-symmetric elastic tensor. Our discussion pertains to plastic responses that are accompanied by emergent topological charges in the response, rather than in the underlying structure of the material. We find additional moduli, but they do not appear in the relation between stress and strain, but rather in the relation between quadrupolar and dipolar fields and the displacement field. 

A central aim of this paper was to explore further the consequences of Chiral symmetry breaking in the mechanical response of amorphous solids. This stems from the constitutive relation between the dipole and the displacement that the theory predicts, i.e. 
\begin{equation}
	\B {\C P} =-\B\Gamma \B d \ .
	\label{constit}
\end{equation}
Observing the explicit form Eq.~(\ref{product}),we see that for $\kappa_o=0$ the prediction is that the dipole field should be co-linear with the displacement field. With $\kappa_o\ne 0$ we expect rotation, with an angle between the inverse displacement field and the dipole field as seen in Eq.~\ref{defphi}. This angle can be positive or negative, but given for any particular displacement field in a given system.
This symmetry breaking is additional to the breaking of translational symmetry. The latter is responsible for the introduction of a typical screening length, which in the notation of the present paper is $\kappa_e^{-1}$. Translational symmetry breaking due to dipole screening is also the mechanism behind the hexatic phase transition \cite{78HN} and the Kosterlitz-Thouless transition \cite{16Kos}. The appearance of this Chiral symmetry breaking seems novel to the mechanical response of amorphous solids, and it was argued in Ref. \cite{24CSWDM} that it is related to irreversibility and energy loss (or gain) in a closed loop of straining an amorphous solids.  
It is worthwhile to remark that the similar notion of ``Odd Elasticity" that was discussed above was been coined in the context of active materials whose mechanics is not derived from a potential \cite{21BSVV}. The present formulation thus may be referred to as ``Odd Plasticity", as it also connects to the lack of energy conservation.   

The methods employed above translate naturally to experimental data. All that one needs is an accurate map of the displacement field, something that was shown to be very doable in experiments using transparent plastic disks, see for example Ref.~\cite{22MMPRSZ}. Once the displacement field is given, the protocol followed in the previous section is relatively straightforward. It seems quite obvious that the interest in the presence of Chiral symmetry breaking warrants a serious experimental attempt to corroborate the simulation results shown above.

An issue that was not explicitly discussed in the present paper is the transition from elastic response to anomalous one. It was shown before that given glass formers or granular system can show both behaviors with an interesting transition between them, as a function of a control parameter like pressure, \cite{24JPS}. We now expect that the transition will be accompanied by both translation and Chiral symmetry breaking, requiring additional scrutiny of the nature of this transition. This is work in progress in our group that will be reported in the near future.

Finally, dynamical response to oscillatory or any form of dynamical straining are known to show dipole screening as well, cf. Ref.~\cite{24HPPS}. The presence of Chiral symmetry breaking and it consequences needs to be examined for such protocols as well. 

\acknowledgments
This work had been supported in part by ISF under grant \#3492/21 (collaboration with China) and the Minerva Center for ``Aging, from physical materials to human tissues" at the Weizmann Institute.

\appendix

\section{Analytic solutions of the equations}
\label{anal}

In this appendix we solve the Eqs.~(\ref{L4}) and (\ref{L5}) for the analytical form of the radial and transverse displacements functions $d_{r}$ and $d_{\theta}$ respectively.  
We consider the following forms of the $d_{r}$ and $d_{\theta}$,
\begin{align}\label{A0}
	&d_{r}(r,\theta) = a_{0}(r)  \nonumber \\
	&d_{\theta}(r,\theta)  = b_{0}(r)  .
\end{align}
Thus equations (\ref{L4} and \ref{L5}) reduce to following coupled equations in $a_{0}(r)$ and $b_{0}(r)$,
\begin{align}\label{A00}
	[r^2a_0^{\prime\prime} + ra_0^{\prime}-a_0 ]  +  \frac{\kappa_{e}^{2}r^2}{(\tilde\lambda +2)}  a_0 -\frac{\kappa_o^{2}r^2}{(\tilde\lambda +2)}  b_0   & =0 ,  \nonumber \\ 
	[r^2b_0^{\prime\prime}  + rb_0^{\prime} -b_0 ]  + \kappa_{e}^{2} r^2 b_0  +   \kappa_o^{2} r^{2} a_0    &=0.
\end{align}
We can combine the above two coupled equations in $a_{0}(r)$ and $b_{0}(r)$ by using two Lagrange multipliers $C_1$ and $C_2$ as follows
\begin{widetext}
	\begin{align}\label{A1}
		C_1 \left\lbrace [r^2a_0^{\prime\prime} + ra_0^{\prime}-a_0 ]  +  \frac{\kappa_{e}^{2}r^2}{(\tilde\lambda +2)}  a_0 -\frac{\kappa_o^{2}r^2}{(\tilde\lambda +2)}  b_0  \right\rbrace   
		+ C_2 \left\lbrace [r^2b_0^{\prime\prime}  + rb_0^{\prime} -b_0 ]  + \kappa_{e}^{2} r^2 b_0  +   \kappa_o^{2} r^{2} a_0 \right\rbrace  =0, 
	\end{align}
	which after some simplification can be written as,
	\begin{align}\label{A2}
		r^2\left[ C_1 a_0^{\prime\prime} + C_2 b_0^{\prime\prime}\right]  +   r\left[ C_1 a_0^{\prime} + C_2 b_0^{\prime}\right]  -  \left[ C_1 a_0 + C_2 b_0\right]  + \left[ C_1 \frac{\kappa_{e}^{2}r^2}{(\tilde\lambda +2)} + C_2 \kappa_o^{2} r^{2} \right] a_0 + \left[  C_2 \kappa_{e}^{2} r^2  -  C_1 \frac{\kappa_o^{2}r^2}{(\tilde\lambda +2)}   \right] b_0 =0 .
	\end{align}
\end{widetext}
The above equation can be written in the form of Bessel differential equation. To show that this is true, we write the above equation in the following form
\begin{multline}\label{A3}
	r^2\left[ C_1 a_0^{\prime\prime} + C_2 b_0^{\prime\prime}\right]  +   r\left[ C_1 a_0^{\prime} + C_2 b_0^{\prime}\right]  -  \left[ C_1 a_0 + C_2 b_0\right] \\
	+ r^2[\tilde{C_1} a_0 + \tilde{C_2} b_0] =0 ,
\end{multline}
where,
\begin{align}\label{A4}
	& \tilde{C_1} = \left[ C_1 \frac{\kappa_{e}^{2}}{(\tilde\lambda +2)} + C_2\kappa_o^{2} \right], \nonumber \\
	& \tilde{C_2} = \left[  C_2 \kappa_{e}^{2}  -  C_1 \frac{\kappa_o^{2}}{(\tilde\lambda +2)}   \right].
\end{align}

Now let us substitute 
\begin{align}\label{A5}
	r^2[\tilde{C_1} a_0 + \tilde{C_2} b_0] =  \kappa^2 r^2[ C_1  a_0 +  C_2  b_0],
\end{align}
where $\kappa$ is a parameter to be obtained such that it defines the above transformation. Thus
\begin{align}\label{A6}
	[\tilde{C_1} a_0 + \tilde{C_2} b_0] =   [  \kappa^2 C_1  a_0 +  \kappa^2 C_2  b_0],
\end{align}
or,
\begin{align}\label{A7}
	[\tilde{C_1} -  \kappa^2 C_1 ]a_0 + [\tilde{C_2} -\kappa^2 C_2 ]b_0 =0 .
\end{align}
Since, $a_0(r)$ and $b_0(r)$ are arbitrary, therefore to hold the above equation true, we must have 
\begin{align}\label{A8}
	& \tilde{C_1} =  \kappa^2 C_1 , \nonumber \\
	& \tilde{C_2} =   \kappa^2 C_2.
\end{align}
Now we can solve these two equations together to determine the value of $\kappa$. After substituting the values of $ \tilde{C_1}$  and $ \tilde{C_2}$, above equations reduce to
\begin{align}\label{A9}
	&  C_1 \frac{\kappa_{e}^{2}}{(\tilde\lambda +2)} + C_2 \kappa_o^{2}  =  \kappa^2 C_1, \nonumber \\
	&    C_2 \kappa_{e}^{2}  -  C_1 \frac{\kappa_o^{2}}{(\tilde\lambda +2)}  =   \kappa^2 C_2.
\end{align}
With further simplifications we have
\begin{align}\label{A10}
	&  \frac{\kappa_{e}^{2}}{(\tilde\lambda +2)} + \frac{C_2}{C_1} \kappa_o^{2}   =  \kappa^2 , \nonumber \\ \nonumber \\
	&   \kappa_{e}^{2}  - \frac{C_1}{C_2} \frac{\kappa_o^{2}}{(\tilde\lambda +2)}  =   \kappa^2 .
\end{align}
Let us take $\frac{C_1}{C_2} = Z$, then the above equations simplify to
\begin{align}\label{A11}
	&  \frac{\kappa_{e}^{2}}{(\tilde\lambda +2)} + \frac{1}{Z} \kappa_o^{2}   =  \kappa^2 , \nonumber \\ \nonumber \\
	&   \kappa_{e}^{2}  - Z \frac{\kappa_o^{2}}{(\tilde\lambda +2)}  =   \kappa^2 .
\end{align}
Now we have two equations giving the same values of $\kappa$. We equate these two equation to give a quadratic equation in $Z$,
\begin{align}\label{A12}
	\frac{\kappa_{e}^{2}}{(\tilde\lambda +2)} + \frac{1}{Z} \kappa_o^{2}   =    \kappa_{e}^{2}  - Z \frac{\kappa_o^{2}}{(\tilde\lambda +2)}.   
\end{align}

Which simplifies to
\begin{align}\label{A13}
	Z^2 \frac{\kappa_o^{2}}{ \tilde\lambda +2 } + \left(  \frac{\kappa_{e}^{2}}{ \tilde\lambda +2 }  -  \kappa_{e}^{2}\right)  Z +  \kappa_o^{2}  =0,
\end{align}
or 
\begin{align}\label{A14}
	\kappa_o^{2} Z^2  -\kappa_{e}^2(\tilde\lambda + 1)  Z +   \kappa_o^{2} (\tilde\lambda + 2)  =0.
\end{align}
Now if we substitute $Z=\frac{C_1}{C_2}$, then the above equation can be written in terms of $C_1$ and $C_2$ as follows
\begin{align}\label{A15}
 \kappa_o^{2} C_{1}^2  -\kappa_{e}^2(\tilde\lambda + 1)  C_{1}C_{2} +   \kappa_o^{2} (\tilde\lambda + 2)C_{2}^2  =0.
\end{align}

The solutions of $Z$ are
\begin{equation}\label{A16}
	Z = \frac{\kappa_{e}^2 (\tilde\lambda + 1) \pm \sqrt{\kappa_{e}^4 (\tilde\lambda + 1)^2 - 4(\tilde\lambda + 2)\kappa_o^4 } }{2\kappa_o^2},
\end{equation}
or 
\begin{equation}\label{A17}
	Z_1  = \frac{\kappa_{e}^2 (\tilde\lambda + 1) + \sqrt{\kappa_{e}^4 (\tilde\lambda + 1)^2 - 4(\tilde\lambda + 2)\kappa_o^4 } }{2\kappa_o^2}, 
\end{equation}
and 
\begin{equation}\label{A18}
	Z_2  = \frac{\kappa_{e}^2 (\tilde\lambda + 1) - \sqrt{\kappa_{e}^4 (\tilde\lambda + 1)^2 - 4(\tilde\lambda + 2)\kappa_o^4 } }{2\kappa_o^2}. 
\end{equation}
Substituting the values of $Z$ from above equations into Eqn.(\ref{A11}), we obtain the values of $\kappa$ defining the transformation in Eqn.(\ref{A5}). Therefore, from equations (\ref{A3}, and \ref{A5}) we have
\begin{multline}\label{A19}
	r^2\left[ C_1 a_0^{\prime\prime} + C_2 b_0^{\prime\prime}\right]  +   r\left[ C_1 a_0^{\prime} + C_2 b_0^{\prime}\right]  -  \left[ C_1 a_0 + C_2 b_0\right] \\ + \kappa^2 r^2[C_1 a_0 + C_2 b_0] =0 ,
\end{multline}
or 
\begin{multline}\label{A20}
	r^2\left[ C_1 a_0^{\prime\prime} + C_2 b_0^{\prime\prime}\right]  +   r\left[ C_1 a_0^{\prime} + C_2 b_0^{\prime}\right] \\ + (\kappa^2 r^2 -1)[C_1 a_0 + C_2 b_0] =0.
\end{multline}
We can simplify it little bit more to give 
\begin{multline}\label{A21}
	r^2\left[ \frac{C_1}{C_2} a_0^{\prime\prime} +  b_0^{\prime\prime}\right]  +   r\left[ \frac{C_1}{C_2} a_0^{\prime} +  b_0^{\prime}\right] \\
	+ (\kappa^2 r^2 -1)\left[ \frac{C_1}{C_2} a_0 +  b_0\right]  =0 .
\end{multline}

Now, if we substitute
\begin{align}\label{A22}
	X(r) &= \frac{C_1}{C_2}a_0(r) + b_{0}(r) \nonumber \\
	&=  ~~Z a_0(r) + b_{0}(r),
\end{align}
in the Eqn.(\ref{A21}), we obtain the following differential equation
\begin{align}\label{A23}
	r^2X^{\prime\prime} + rX^{\prime}  + (\kappa^2 r^2 -1)X =0.
\end{align}
This is a bessel differential equation, where $\kappa$ and $Z$ are already defined above. A general solution of this equation is
\begin{align}\label{A24}
	X(r) = m J_{1}(\kappa r) + n Y_{1}(\kappa r),
\end{align}
where $J_{1}$ and $Y_{1}$ are the Bessel functions of first kind, and the coefficients $m$ and $n$ are the constant parameters to be obtained using the boundary conditions.
Note that, we will have two solutions corresponding to the two values of $\kappa$ or $Z$.
The two values of $Z$ are $Z_1$ and $Z_2$, and the corresponding values of $\kappa$ are $\eta$ and $\zeta$ respectively.
Then we obtain following two coupled equations in $a_{0}(r)$ and $b_{0}(r)$ from the equations (\ref{A22}), and (\ref{A24}),
\begin{align}\label{A25}
	m_1 J_{1}(\eta r) + n_1 Y_{1}(\eta r) = Z_1 a_0(r) + b_0(r),
\end{align}
and
\begin{align}\label{A26}
	m_2 J_{1}(\zeta r) + n_2 Y_{1}(\zeta r) = Z_2 a_0(r) + b_0(r).
\end{align}
From the above two equations we immediately obtain the analytical forms for $a_0(r)$, and $b_0(r)$
\begin{widetext}
	\begin{align}\label{good1}
		a_{0}(r) = \frac{ \left[ [ m_1 J_{1}(\eta r) + n_1 Y_{1}(\eta r)] - [m_2 J_{1}(\zeta r) + n_2 Y_{1}(\zeta r)] \right] }{ Z_1 -Z_2  },
	\end{align}
	\begin{align}\label{good2}
		b_{0}(r) = \frac{ Z_2 [ m_1 J_{1}(\eta r) +  n_1 Y_{1}(\eta r)] - Z_1 [ m_2 J_{1}(\zeta r) + n_2 Y_{1}(\zeta r) ] } { \left(  Z_2 - Z_1 \right)  }.
	\end{align}
\end{widetext}
Now we use boundary conditions to determine the coefficients $m_1, n_1$ and $m_2, n_2$. 
The boundary conditions on $a_0(r)$ and $b_0(r)$ are 
\begin{align}\label{A29}
	a_{0}(r)|_{r=r_{in}} = d_{a},  ~~~~~~~~~~~~ a_{0}(r)|_{r=r_{o}} = 0, \nonumber \\
	b_{0}(r)|_{r=r_{in}} = d_{b},  ~~~~~~~~~~~~ b_{0}(r)|_{r=r_{o}} = 0.
\end{align}
With the above boundary conditions, we have following four coupled equations in $m_1, n_1$ and $m_2, n_2$, 
\begin{widetext}
	\begin{align}\label{A30}
		[ m_1 J_{1}(\eta r_{in}) + n_1 Y_{1}(\eta r_{in})] - [m_2 J_{1}(\zeta r_{in}) + n_2 Y_{1}(\zeta r_{in})]  &= (Z_1 -Z_2)d_a, ~~~~~~~~~~~~~~(i)\nonumber \\ \nonumber \\
		[ m_1 J_{1}(\eta r_{o}) + n_1 Y_{1}(\eta r_{o})] - [m_2 J_{1}(\zeta r_{o}) + n_2 Y_{1}(\zeta r_{o})]   &=0, ~~~~~~~~~~~~~~~~~~~~~~~~~~~~(ii) \nonumber  \\   \\
		Z_2 [ m_1 J_{1}(\eta r_{in}) +  n_1 Y_{1}(\eta r_{in})] - Z_1 [ m_2 J_{1}(\zeta r_{in}) + n_2 Y_{1}(\zeta r_{in}) ] &= (Z_2 -Z_1)d_b, ~~~~~~~~~~~~~~(iii)\nonumber \\ \nonumber \\
		Z_2 [ m_1 J_{1}(\eta r_{o}) +  n_1 Y_{1}(\eta r_{o})] - Z_1 [ m_2 J_{1}(\zeta r_{o}) + n_2 Y_{1}(\zeta r_{o}) ] &= 0.  ~~~~~~~~~~~~~~~~~~~~~~~~~~~~(iv)\nonumber 
	\end{align}
\end{widetext}
From Eqn. \ref{A30}$(ii)$, we have 
\begin{align}\label{A31}
	[ m_1 J_{1}(\eta r_{o}) + n_1 Y_{1}(\eta r_{o})] = [m_2 J_{1}(\zeta r_{o}) + n_2 Y_{1}(\zeta r_{o})],
\end{align}
and using this in equation \ref{A30}$(iv)$, we get 
\begin{align}\label{A32}
	[ m_2 J_{1}(\zeta r_{o}) + n_2 Y_{1}(\zeta r_{o}) ] (Z_2 - Z_1) = 0 .
\end{align}
Since $Z_1 \neq Z_2$, therefore $ (Z_2 - Z_1) \neq 0$. Thus we must have
\begin{align}\label{A33}
	m_2 J_{1}(\zeta r_{o}) + n_2 Y_{1}(\zeta r_{o})  =0,
\end{align} 
which gives
\begin{align}\label{A34}
	m_2 = -n_2 \frac{Y_{1}(\zeta r_{o})}{J_{1}(\zeta r_{o})}.
\end{align}
Similarly from Eqns. \ref{A30}$(ii)$, and \ref{A30}$(iv)$, we have 
\begin{align}\label{A35}
	[ m_1 J_{1}(\eta r_{o}) + n_1 Y_{1}(\eta r_{o})](Z_2 -Z_1)  =0, 
\end{align}
which gives
\begin{align}\label{A36}
	m_1 = -n_1 \frac{Y_{1}(\eta r_{o})}{J_{1}(\eta r_{o})}.
\end{align}
\begin{widetext}
	Substituting the values of $m_1$ and $m_2$ into Eqn. \ref{A30}$(i)$, we obtain,
	\begin{align}\label{A37}
		n_{1}\left[   \frac{Y_{1}(\eta r_{in})J_{1}(\eta r_{o}) -  Y_{1}(\eta r_{o})J_{1}(\eta r_{in}) }{J_{1}(\eta r_{o})}  \right]  - n_{2}\left[ \frac{Y_{1}(\zeta r_{in})J_{1}(\zeta r_{o}) -  Y_{1}(\zeta r_{o})J_{1}(\zeta r_{in}) }{J_{1}(\zeta r_{o})}  \right]  = (Z_1 - Z_2)d_a .
	\end{align}
	Similarly substituting the values of $m_1$ and $m_2$ into Eqn. \ref{A30}$(iii)$, we obtain,  
	\begin{align}\label{A38}
		n_{1}Z_2\left[   \frac{Y_{1}(\eta r_{in})J_{1}(\eta r_{o}) -  Y_{1}(\eta r_{o})J_{1}(\eta r_{in}) }{J_{1}(\eta r_{o})}  \right]  - n_{2}Z_1 \left[ \frac{Y_{1}(\zeta r_{in})J_{1}(\zeta r_{o}) -  Y_{1}(\zeta r_{o})J_{1}(\zeta r_{in}) }{J_{1}(\zeta r_{o})}   \right]  = (Z_2 - Z_1)d_b.
	\end{align}
\end{widetext}
For simplicity let us define 
\begin{align}\label{A39}
	\alpha = \left[   \frac{Y_{1}(\eta r_{in})J_{1}(\eta r_{o}) -  Y_{1}(\eta r_{o})J_{1}(\eta r_{in})    }{J_{1}(\eta r_{o})}  \right], \nonumber \\ 
\end{align}
and 
\begin{align}\label{A40}
	\beta =  \left[ \frac{Y_{1}(\zeta r_{in})J_{1}(\zeta r_{o}) -  Y_{1}(\zeta r_{o})J_{1}(\zeta r_{in}) }{J_{1}(\zeta r_{o})}   \right]. \nonumber \\ 
\end{align}
Thus Eqns. (\ref{A37} and \ref{A38}) reduce to
\begin{align}\label{A41}
	n_{1}\alpha  - n_{2}\beta  = (Z_1 - Z_2)d_a ,
\end{align}
and, 
\begin{align}\label{A42}
	n_{1}\alpha Z_2 - n_{2}\beta Z_1 = (Z_2 - Z_1)d_b .
\end{align}
The above two equations can be solved together to give values of $n_1$ and $n_2$
\begin{align}\label{A43}
	n_1 &= \frac{Z_1 d_a + d_b}{\alpha} \nonumber \\
	&= \frac{(Z_1 d_a + d_b)J_{1}(\eta r_{o})}{ [ Y_{1}(\eta r_{in})J_{1}(\eta r_{o}) -  Y_{1}(\eta r_{o})J_{1}(\eta r_{in}) ] } ,
\end{align}
and
\begin{align}\label{A44}
	n_2 &= \frac{Z_2 d_a + d_b}{\beta} \nonumber \\
	& = \frac{(Z_2 d_a + d_b)J_{1}(\zeta r_{o})}{ [ Y_{1}(\zeta r_{in})J_{1}(\zeta r_{o}) -  Y_{1}(\zeta r_{o})J_{1}(\zeta r_{in}) ] }.
\end{align}
Now we can write the values of $m_1$ and $m_2$ from Eqs.(\ref{A34} and \ref{A36})
\begin{align}\label{A45}
	& m_1 =  -\frac{(Z_1 d_a + d_b)Y_{1}(\eta r_{o})}{ [ Y_{1}(\eta r_{in})J_{1}(\eta r_{o}) -  Y_{1}(\eta r_{o})J_{1}(\eta r_{in}) ] }, \nonumber \\ \nonumber \\
	& m_2 = -\frac{(Z_2 d_a + d_b)Y_{1}(\zeta r_{o})}{ [ Y_{1}(\zeta r_{in})J_{1}(\zeta r_{o}) -  Y_{1}(\zeta r_{o})J_{1}(\zeta r_{in}) ] } .
\end{align}
We list below all the parameters that are needed to determine the functional forms of $a_0(r)$ and $b_0(r)$. The values of $Z$ are
\begin{align}\label{A46}
	& Z_1  =  \frac{\kappa_{e}^2 (\tilde\lambda + 1) + \sqrt{\kappa_{e}^4 (\tilde\lambda + 1)^2 - 4(\tilde\lambda + 2)\kappa_o^4 } }{2\kappa_o^2}, \nonumber \\ \nonumber \\
	& Z_2  = \frac{\kappa_{e}^2 (\tilde\lambda + 1) - \sqrt{\kappa_{e}^4 (\tilde\lambda + 1)^2 - 4(\tilde\lambda + 2)\kappa_o^4 } }{2\kappa_o^2}. 
\end{align}
The two values of $\kappa$, i.e., $\eta$ and $\zeta$ are 
\begin{align}\label{A47}
	&  \eta = \sqrt{ \kappa_{e}^{2}  - Z_1 \frac{\kappa_o^{2}}{(\tilde\lambda +2)}}, \nonumber \\ \nonumber \\
	&  \zeta = \sqrt{ \kappa_{e}^{2}  - Z_2 \frac{\kappa_o^{2}}{(\tilde\lambda +2)}}.
\end{align}
and the parameters $m_1, n_1, m_2, n_2$ are
\begin{align}\label{A48}
	& n_1 =  \frac{(Z_1 d_a + d_b)J_{1}(\eta r_{o})}{ [ Y_{1}(\eta r_{in})J_{1}(\eta r_{o}) -  Y_{1}(\eta r_{o})J_{1}(\eta r_{in}) ] }, \nonumber \\ \nonumber \\
	& m_1 =  -\frac{(Z_1 d_a + d_b)Y_{1}(\eta r_{o})}{ [ Y_{1}(\eta r_{in})J_{1}(\eta r_{o}) -  Y_{1}(\eta r_{o})J_{1}(\eta r_{in}) ] }, \nonumber \\ \nonumber \\
	& n_2 =  \frac{(Z_2 d_a + d_b)J_{1}(\zeta r_{o})}{ [ Y_{1}(\zeta r_{in})J_{1}(\zeta r_{o}) -  Y_{1}(\zeta r_{o})J_{1}(\zeta r_{in}) ] },\nonumber \\ \nonumber \\
	& m_2 = -\frac{(Z_2 d_a + d_b)Y_{1}(\zeta r_{o})}{ [ Y_{1}(\zeta r_{in})J_{1}(\zeta r_{o}) -  Y_{1}(\zeta r_{o})J_{1}(\zeta r_{in}) ] } .
\end{align}
\section{protocols of the preparation of the amorphous solid}
\label{numerics}

Here we outline the details of the numerical simulation to produce the glassy configurations. We study a two-dimensional poly-dispersed model of $N=20,000$ point particles in an annulus bounded by two rigid walls, with an inner radius, $r_{in}$ and outer radius, $r_{out}$. The particles are filled in annular region such that density of the system, $\rho=N/A=1.0$, where $A = \pi (r_{out}^2 -r_{in}^2)$. The binary interactions between point particles with mass, $m=1$ is given Lennard-Jones (LJ) potential:

\begin{equation}
	u_{ij}(r)=
	\begin{cases}
		u_{ij}^{LJ} + A_{ij}+B_{ij} r + C_{ij}r^2, \;\;\textbf{if}\; R \leq R_{ij}^{cut} \\
		0, \hskip 30mm\text {otherwise}
	\end{cases}
	\label{eq1}
\end{equation}
where
\begin{equation}
	u_{ij}^{LJ}= 4\epsilon_{ij} \Big[ \Big(\frac{\sigma_{ij}}{r}\Big)^{12} - \Big( \frac{\sigma_{ij}} {r} \Big)^6  \Big]
\end{equation}

\noindent The parameters $A_{ij} = 0.4526\epsilon_{ij}$, $B = -0.3100\epsilon_{ij}/\sigma_{ij}$, $C = 0.0542\epsilon_{ij}/\sigma_{ij}^2$ are added to smooth the potential, $u_{ij}(r)$ at cut-off distance, $R_{ij}^{cut} = 2.5\sigma_{ij}$ (upto second derivative). The interaction length of particles, $\sigma_i$ is drawn from a  probability distribution, $P(\sigma) \simeq 1/\sigma^3$ in a range between $\sigma_{max}$ and $\sigma_{min}$ such that the mean $\bar{\sigma}=1$. The mixing rule of $\sigma_i$ for binary interaction are:
\begin{equation}
	\begin{split}
		\sigma_{ij} = \frac{\sigma_i + \sigma_j}{2} \Big[1-0.2|\sigma_i - \sigma_j| \Big],\\
		\sigma_{max} = 1.61, \sigma_{min}=\sigma_{max} /2.219
	\end{split}
\end{equation}

The reduced units for mass, length, energy and time are $m$, $\bar{\sigma}$, $\epsilon_{ij}=1$ and $\bar{\sigma}\sqrt{m/\epsilon_{ij}}$ respectively. The interaction between point particles and the walls is of same form as Eq \ref{eq1}, where $r_{ij}$ is replaced by distance to the wall.

The system is first thermalized at \textquotedblleft mother temperature\textquotedblright, $T_m=1$  using swap Monte Carlo and cooled down to $T=0$ using conjugate gradient method. Once the system is mechanically equilibrated with total force on each particle smaller than $10^{-8}$, we inflate the inner radius $r_{in}$ such that inner radius after inflation is $r_{in}+d_0$.

After inflation, we equilibrate the system again using conjugate gradient method and measure the displacement field, $\boldsymbol{d}$ by comparing the equilibrated configurations before and after inflation. To compare with the theory, we compute the angle-averaged radial component of the radial displacement field, $d_r(r)$.

\bibliography{ALL.anomalous}

\end{document}